\documentclass[11pt,a4paper]{article}
\usepackage{jcappub}

\usepackage{amsmath,amssymb,bm,float,mathrsfs,subfigure,tabularx}
\usepackage{graphicx}
\usepackage{dcolumn}
\usepackage{color}
\usepackage{hyperref}
\usepackage{comment}

  \newcommand{\beq}{\begin{equation}}
  \newcommand{\eeq}{\end{equation}}
  \newcommand{\al}[1]{\begin{align} #1 \end{align}}
  \newcommand{\bi}{\begin{itemize}}
  \newcommand{\ei}{\end{itemize}}
  \newcommand{\bc}{\begin{center}}
  \newcommand{\ec}{\end{center}}
  
  \def\dd{\mathrm{d}}
  \def\DD{\mathrm{D}}

  \def\rmE{\mathrm{E}}
  \def\rmJ{\mathrm{J}}
  \def\rmL{\mathrm{L}}
  
  \def\rmR{\mathrm{R}}

  \def\mcA{\mathcal{A}}

  \def\mcK{\mathcal{K}}
  
  \def\mcM{\mathcal{M}}
  \def\mcN{\mathcal{N}}
  \def\mcO{\mathcal{O}}
  \def\mcP{\mathcal{P}}

  \def\mcS{\mathcal{S}}

  \def\mcY{\mathcal{Y}}
  
  \def\pd{\partial}

  \newcommand{\ave}[1]{\left\langle #1 \right\rangle}

  \def\3Dint#1{\int\frac{\dd^{3}{#1 }}{(2\pi )^3}}

{\baselineskip0pt
\rightline{\baselineskip16pt\rm\vbox to20pt{
            \hbox{RESCEU-2/14, IPMU14-0031}
\vss}}%
}

\title{Is there supercurvature mode of massive vector field in open
inflation?}

\author[a]{Daisuke Yamauchi}
\author[b,c]{Tomohiro Fujita}
\author[b]{Shinji Mukohyama}

\affiliation[a]{%
Research Center for the Early Universe, Graduate School of Science, 
The University of Tokyo, Bunkyo-ku, Tokyo 113-0033, Japan
}%

\affiliation[b]{%
Kavli Institute for the Physics and Mathematics of the Universe (WPI), TODIAS, The University 
of Tokyo, Kashiwa, Chiba 277-8568, Japan 
}%

\affiliation[c]{%
Department of Physics, The University of Tokyo,
Bunkyo-ku, Tokyo 113-0033, Japan
}%

\emailAdd{yamauchi@resceu.s.u-tokyo.ac.jp}
\emailAdd{tomohiro.fujita@ipmu.jp}
\emailAdd{shinji.mukohyama@ipmu.jp}

\abstract{
We investigate the Euclidean vacuum mode functions of a massive vector
field in a spatially open chart of de Sitter spacetime. In the
one-bubble open inflationary scenario that naturally predicts a negative 
spatial curvature after a quantum tunneling, it is known that a light
scalar field has the so-called supercurvature mode, i.e. an additional
discrete mode which describes fluctuations over scales larger than the
spatial curvature scale. If such supercurvature modes exist for a vector
field with a sufficiently light mass, then they would decay slower and 
easily survive the inflationary era. However, the existence of
supercurvature mode strongly depends on details of the system. To
clarify whether a massive vector field has supercurvature modes, we
consider a U(1) gauge field with gauge and conformal invariances 
spontaneously broken through the Higgs mechanism, and present 
explicit expressions for the Euclidean vacuum mode functions. We find
that, for any values of the vector field mass, there is no
supercurvature mode. In the massless limit, the absence of
supercurvature modes in the scalar sector stems from the gauge
symmetry. 
}

\begin{document}

 \maketitle

\section{Introduction} 

Recent observational data provide a strong support of the existence of 
extragalactic magnetic fields, in the range of $\mcO (10^{-14}$-$10^{-20})\,{\rm G}$
on Mpc scales~\cite{Neronov:1900zz,Tavecchio:2010mk,Dermer:2010mm,Huan:2011kp,Dolag:2010ni,Essey:2010nd,Taylor:2011bn,Vovk:2011aa,Takahashi:2011ac,Finke:2013bua}\,.
The generation of the magnetic field in high-redshift galaxies, clusters, and 
even in empty intergalactic region is still an unresolved problem in cosmology.
No promising astrophysical process to generate the sufficient amount of
the magnetic field on the large scales are known.
As for the inflationary magnetogenesis, though the various mechanism are proposed,
the several difficulties such as the strong coupling problem, the backreaction problem and the curvature perturbation problem
in some specific models prevent successful production of magnetic field~\cite{Demozzi:2009fu, Barnaby:2012xt, Fujita:2013qxa}.
Actually both upper and lower limits on the inflation energy scale
can be derived from these problems
in model independent ways
and the limits are considerably severe if the extragalactic magnetic fields
are stronger than $10^{-16}$G at present~\cite{Fujita:2012rb,Suyama:2012wh,Fujita:2014sna}.
Thus it is known to be very difficult to generate the magnetic field in the context of the inflationary magnetogenesis
on the flat Friedmann-Lema\^itre-Robertson-Walker (FLRW) universe.

The superadiabatic growth of the magnetic fields in the open FLRW
universe has been discussed in the literatures~\cite{Barrow:2011ic,Barrow:2012ty}.
The authors of these literatures assumed the existence of supercurvature
modes of the magnetic field, which describes the fluctuations with the
wavelength exceeding the spatial curvature scale. If a supercurvature
mode exists, it decays slower than $1/a^2$\,, where $a$ corresponds to
the conventional scale factor of a FLRW universe, and can easily survive
the inflationary era. Hence the relatively large amount of the magnetic
field on supercurvature scales would remain at late time. However, the 
existence of supercurvature modes of magnetic fields is non-trivial and 
should be critically studied. Adamek {\it et al.}~\cite{Adamek:2011hi}
recently pointed out that the equations of motion of a U(1) gauge field
with unbroken conformal and gauge symmetries can be rewritten in the
form that is identical to those of massive scalar fields for which
there is no supercurvature mode~\footnote{Rigorously speaking, the proof
of the absence of supercurvature modes requires knowledge of not only
the equation of motion but also a Klein-Gordon norm and proper boundary
conditions. The present paper fills those gaps for the analysis of the
massless vector field, although our main focus will be on a massive
vector field.}.

The purpose of the present paper is to investigate whether 
supercurvature modes exist for a massive vector field, in both scalar
and vector sectors of the physical spectrum. To be specific, we consider
a U(1) gauge field with both gauge and conformal symmetries
spontaneously broken through the Higgs mechanism. As for the background
geometry, we consider a de Sitter spacetime in the open chart. This is
relevant to the one-bubble open inflation scenario that naturally
predicts the spatially negative curvature universe. While the recent
observational data show that the universe is almost exactly flat with
accuracy of about $1\%$\,, 
$|1-\Omega_0 |\leq 10^{-2}$~\cite{Ade:2013zuv}\,,
open inflation scenario is attracting a renewed interest in the context
of the string landscape
scenario~\cite{Susskind:2003kw,Freivogel:2004rd}. 
There are a huge number of metastable de Sitter vacua and the tunneling
transition generally occurs through the nucleation of a true vacuum
bubble in the false vacuum background. Because of the symmetry of the
instanton solution, a bubble formed by the Coleman-De Luccia (CDL)
instanton~\cite{Coleman:1977py,Coleman:1980aw} looks like an infinite
open universe from the viewpoint of an observer inside. If the universe
experienced a sufficiently long inflation after the bubble nucleation,
then the universe becomes almost exactly flat and subsequently evolves
as a slightly open FLRW universe. This leads to a natural realization of
one-bubble open inflation (see
e.g. \cite{Sasaki:1994yt,Yamamoto:1996qq,Tanaka:1997kq,Garriga:1998he,Garriga:1997wz})
and can be tested against
observations~\cite{Yamauchi:2011qq,Sugimura:2012kr,Sugimura:2013cra}.

This paper is organized as follows. We first illustrate the background
spacetime in section \ref{sec:Background}. In section 
\ref{sec: KG norm}\,, we expand the U(1) gauge field by harmonic
functions and write down the reduced action for the even and odd modes
of the U(1) gauge field. In order to investigate the existence/absence
of supercurvature modes, we show the quantization conditions for the
even and odd modes on a Cauchy surface. With the obtained normalization
conditions, we then analyze whether the supercurvature modes, which are
normalizable on the Cauchy surface, exist in section 
\ref{sec: Mode functions}.  In section \ref{sec: massless limit}\,, as a
consistency check, we explicitly calculate the Wightman function in the
decoupling limit by using the (subcurvature) mode functions derived in
section \ref{sec: KG norm}. It is shown that the correct expression for
the Euclidean Wightman function is recovered in the decoupling limit
without need for any supercurvature modes. Finally, section
\ref{sec:summary} is devoted to a summary and discussions.

\section{Background}
\label{sec:Background}

In this paper, we consider a U(1) gauge field with both gauge and
conformal symmetries spontaneously broken through the Higgs mechanism in
an open de Sitter geometry, i.e. a de Sitter spacetime in the open
chart.

Before showing relevant forms of the background metric, we illustrate
that this setup is appropriate to investigate the existence/absence of 
supercurvature mode of a massive vector field in open inflationary
universe. Let us begin with a system which consists of multi scalar
fields and the U(1) gauge field minimally coupled with Einstein
gravity. We investigate the evolution of mode functions of the U(1)
gauge field in the one-bubble open inflationary scenario and
particularly focus on whether the supercurvature modes are generated. 
To be specific, we introduce a real scalar field $\sigma$ that governs
the quantum tunneling from a false vacuum to a true vacuum and realizes
inflation after the quantum tunneling, and a complex scalar field $\Phi$
that plays a major role in the coupling to the U(1) gauge field $A_\mu$. 
Our action is given by
\al{
        S=S_{\rm tun}+S_{\rm AH}
        \,,
}
where
\al{
        &S_{\rm tun}=\int\dd^4 x\sqrt{-g}
                \Biggl[
                        \frac{M_{\rm pl}^2}{2}R
                        -\frac{1}{2}g^{\mu\nu}\pd_\mu\sigma\pd_\nu\sigma 
                        -V_{\rm tun}(\sigma )
                \Biggr]
        \,,\\ 
        &S_{\rm AH}
                =\int\dd^4 x\sqrt{-g}
                        \Biggl[
                                -\frac{1}{4}F_{\mu\nu}F^{\mu\nu}
                                -g^{\mu\nu}\DD_\mu\Phi\overline{\DD_\nu \Phi}
                                -V_{\Phi}(|\Phi |)
                        \Biggr]
        \,.\label{eq: first action}
}
Here the potential $V_{\rm tun}(\sigma)$ is assumed to be the form that
realizes the fast vacuum decay, $\DD_\mu =\pd_\mu - ieA_\mu$ is the
gauge-covariant derivative, 
$F_{\mu\nu} = \partial_{\mu} A_\nu-\partial_\nu A_\mu$ is the field 
strength of the gauge field, and an overbar denotes the complex
conjugate. Since the potential term of $\Phi$ depends only on its
absolute value, this action has the local U(1) symmetry: 
\al{
        \Phi\rightarrow \Phi\, e^{i\alpha (x)}
        \,,\ \ \ 
        A_\mu\rightarrow A_\mu -\frac{1}{e}\pd_\mu\alpha (x)
        \,.\label{eq:local U(1) symmetry}
}
However, if $\Phi$ acquires a non-zero vacuum expectation value, 
$\ave{\Phi} \neq 0$, then the local U(1) symmetry is spontaneously
broken. In that case, the phase degree of freedom is absorbed into the
vector field and the gauge field becomes massive as it is well known as
Higgs mechanism. 
In this paper, we consider a simple open inflation model in which
the bubble nucleation can be well described by the single-field
Coleman-de Luccia (CDL) instanton~\cite{Coleman:1977py,Coleman:1980aw}
on the exact de Sitter spacetime with the Hubble parameter $H$\,. 
Hence we assume that the tunneling transition can be described by
a Euclidean $O(4)$-symmetric bounce solution on a Euclidean de Sitter
geometry.

\bc
\begin{figure}[tbp]
\bc
\includegraphics[width=80mm]{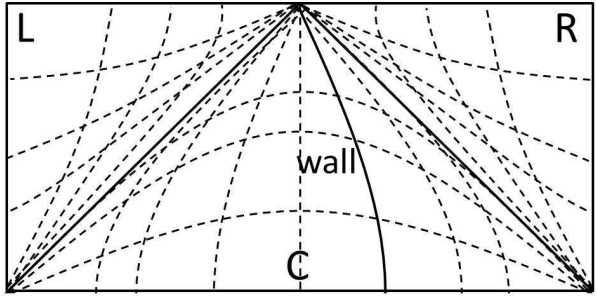}
\caption{
Penrose diagram of bubble nucleating universe.
}
\label{fig:Penrose_diagram}
\ec
\end{figure} 
\ec

The Euclidean geometry can then be well described by the Euclidean de
Sitter metric: 
\al{
        \dd s^2=a_\rmE^2 (\eta_\rmE )\Bigl[\dd\eta_\rmE^2 +\dd r_\rmE^2 +\sin^2 r_\rmE\,\omega_{ab}\dd\theta^a\dd\theta^b\Bigr]
        \,,\label{eq:metric in E}
}
where $-\infty \leq\eta_\rmE\leq +\infty$\,, $0\leq r_\rmE\leq\pi$\,,
$a_\rmE (\eta_\rmE )=1/H\cosh\eta_\rmE$\,, and 
$\omega_{ab}={\rm diag}(1,\sin^2\theta )$ denotes the metric on the unit
$2$-sphere. The background geometry in the Lorentzian regime is obtained
by analytic continuation of the bounce solution. The coordinates in the
Lorentzian regime are 
\al{
        &\eta_\rmE =\eta =-\eta_\rmR -\frac{\pi}{2}i=\eta_\rmL +\frac{\pi}{2}i
        \,,\label{eq:eta relation}\\
        &r_\rmE =-ir +\frac{\pi}{2}=-ir_\rmR =-ir_\rmL
        \,,\label{eq:r relation}\\
        &a_\rmE =a =ia_\rmR =ia_\rmL
        \,.
}
Each set of these coordinates covers one of three distinct parts
of the Lorentzian de Sitter spacetime, called regions-R, L, and C.
Hereafter, we suppress the subscript C because we mainly work in the
region-C. The Penrose diagram for this open FLRW universe is presented in 
Fig.~\ref{fig:Penrose_diagram}. As seen in
Fig.~\ref{fig:Penrose_diagram}, the surfaces which respect the maximal
symmetry in the region-R and L, i.e. $\eta_{\rm R,L} =$const hypersurfaces
are not Cauchy surfaces of the whole spacetime, and hence they are not
appropriate to normalize mode functions (see 
e.g. \cite{Sasaki:1994yt,Yamamoto:1996qq}). 
In the region-C, however, $r =$const hypersurfaces behave as Cauchy
surfaces. Therefore, we need to find the reduced action and properly
construct the Klein-Gordon (KG) norm on a Cauchy surface in the
region-C. The analytic continuation of eq.~\eqref{eq:metric in E} to the
region-C is given by
\al{
        \dd s^2
                =a^2 (\eta )\bar g_{\mu\nu}\dd x^\mu\dd x^\nu
                =a^2 (\eta )
                        \Bigl[
                                \dd\eta^2 -\dd r^2 +\cosh^2 r\,\omega_{ab}\dd\theta^a\dd\theta^b
                        \Bigr]
        \,,\label{eq:metric in region-C}
}
where $a(\eta )=1/H\cosh\eta$\,. Note that in the region-C, 
$\eta={\rm const.}$ hypersurfaces are no longer spacelike, and $r$
instead of $\eta$ plays the role of a time coordinate there.

\section{Reduced action and Klein-Gordon norm}
\label{sec: KG norm}

To describe the Euclidean vacuum state, we need a complete set of mode
functions, which should be properly normalized on a Cauchy surface. In
order to determine whether supercurvature modes of the U(1) gauge field
exist or not, we thus have to construct the KG norm on a Cauchy surface
and to check if the modes can be properly normalized. In this section,
we discuss the quantization of the U(1) gauge field in the open chart of 
the de Sitter spacetime and derive the KG norm on a Cauchy
surface. Since we are interested only in the gauge field, we hereafter
neglect the quantum fluctuations of $\varphi\equiv |\Phi|$ and treat it 
as a non-vanishing constant value by assuming that the mass squared
$V_{\Phi}''(|\Phi|)$ around the potential minimum is large
enough. Hence, it is convenient to decompose the complex scalar field
$\Phi$ into its absolute value and phase as 
$\Phi (x)=\varphi\, e^{i \Theta (x)}$ with $\varphi ={\rm const}$\,. 
Based on the gauge transformation property, 
eq.~\eqref{eq:local U(1) symmetry}, one can construct a gauge-invariant variable.
In the case of the nonvanishing coupling
constant, $e\neq 0$, one possible choice of such variable is 
\al{
        \mcA_\mu\equiv A_\mu -\frac{1}{e}\pd_\mu\Theta
        \,. \label{eq:def-mcAmu}
}
This is chosen as an appropriate variable in the unitary gauge :
$\Theta =0$\,. With these assumptions, the relevant part of the action
\eqref{eq: first action} can be written in terms of the gauge-invariant
variable:
\al{
        S_{\rm eff}
                =-\int\dd^4 x\sqrt{-\bar g}
                        \biggl[
                                \frac{1}{4}\bar g^{\mu\alpha}\bar g^{\nu\beta}
                                \left(\pd_\mu\mcA_\nu -\pd_\nu\mcA_\mu\right)\left(\pd_\alpha\mcA_\beta -\pd_\beta\mcA_\alpha\right)
                                +a^2m_A^2\,\bar g^{\mu\nu}\mcA_\mu\mcA_\nu
                        \biggr]
        \,,\label{eq:relevant action}
}
where $\bar g_{\mu\nu}$ is the conformally transformed metric defined in
eq.~\eqref{eq:metric in region-C} and we have introduced $m_A=e\varphi$
to denote the effective mass of the gauge field.

As we mentioned above, we need to work in the region-C, where the
background configuration is spatially inhomogeneous. Hence we should
expand perturbations in a way that respects the symmetry of the
$2$-sphere rather than that of the $3$-hyperboloid on which harmonics of
various types are defined (see Appendix 
\ref{sec:Harmonics in open universe}). To rewrite the action
\eqref{eq:relevant action} in terms of the $(1+1+2)$ decomposition, let
us decompose $A_\mu$ into the variables $\{A_\eta\,,A_r\,,A_a\}$ with
$a=\theta\,,\phi$. Note that $A_\eta$ and $A_r$ behave as the even
parity modes with respect to the two-dimensional rotation. Since $A_a$
behaves as $2$-vector, we can further decompose it into the even and odd
parity parts as 
\al{
        A_a =A^{({\rm e})}_{:a}+\epsilon_a{}^bA_{:b}^{({\rm o})}
        \,,\label{eq:even odd decomposition}
}
where we have introduced the colon ( $:$ ) as the covariant derivative
with respect to the unit $2$-sphere metric $\omega_{ab}$\,, and
$\epsilon^a{}_b$ is the unit anti-symmetric tensor on the unit
$2$-sphere, which are defined in 
eqs.~\eqref{eq:intrinsic covariant derivative}, 
\eqref{eq:antisymmetric tensor}, respectively.

Before constructing the reduced action, let us consider the boundary
conditions at $\eta\rightarrow\pm\infty$ for the gauge-invariant
variables. Since the boundary of the open slice of the de Sitter
spacetime, that is $\eta\rightarrow\pm\infty$\,, is regular, we can
impose the condition that scalar gauge-invariant quantities such as
$F_{\mu\nu}F^{\mu\nu}$ be all regular at $\eta\rightarrow \pm\infty$\,. 
In the case of the nonvanishing coupling constant, we can construct 
another gauge-invariant quantity $\mcA_\mu$ as defined in
(\ref{eq:def-mcAmu}). We can then impose the condition that the tetrad 
components of the gauge-invariant vector, 
$\mcA_\mu e^\mu_{(\alpha )}=(\mcA_\eta /a\,,\mcA_r/a\,,\mcA_a/a)$\,,
be regular at $\eta\rightarrow\pm\infty$\,. Here, 
$\{e^\mu_{(\alpha )}\}$ is a tetrad basis of the de Sitter spacetime. In
consequence, $\mcA_\eta$\,,$\mcA_r$\,, and $\mcA_a$ have to decay as
fast as (or faster than) $e^{-|\eta |}$ at
$\eta\rightarrow\pm\infty$\,.

Hereafter, using the U(1) gauge degree of freedom, we adopt the unitary gauge :
$\Theta =0$\,. The gauge-invariant vector $\mcA_\mu$ then reduces to the
original gauge field $A_\mu$.

We now expand the perturbations in terms of the spherical harmonics as 
\al{
        &A_\eta (\eta ,r,\Omega )
                =\sum_{\ell m}A_\eta^{\ell m}(\eta ,r)Y_{\ell m}(\Omega )
        \,,\ \  
        A_r (\eta ,r,\Omega )
                =\sum_{\ell m}A_r^{\ell m}(\eta ,r)Y_{\ell m}(\Omega )
        \,,\label{eq:spherical expansion 1}\\
        &A^{(\lambda )}(\eta ,r,\Omega )
                =\sum_{\ell m}A^{(\lambda )\ell m}(\eta ,r)Y_{\ell m}(\Omega )
        \,,
        \label{eq:spherical expansion 2}
}
where $\lambda ={\rm e}$ and ${\rm o}$\,. 
It is straightforward to express the action in terms of the coefficients
of the spherical harmonic expansion. We then find that the action can be
decomposed into the even and odd parity parts as
\al{
        S_{\rm eff}=S^{({\rm e})}+S^{({\rm o})}
        \,,\label{eq:reduced action}
}
where
\al{
        &S^{({\rm e})}
                =\frac{1}{2}\sum_{\ell m}\int\dd r\dd\eta
                        \biggl\{
                                \cosh^2 r\left(\pd_r A_\eta^{\ell m} -\pd_\eta A_r^{\ell m}\right)^2
                                +a^2m_A^2\cosh^2 r
                                        \biggl[
                                                \left( A_r^{\ell m}\right)^2 -\left( A_\eta^{\ell m}\right)^2
                                        \biggr]
        \notag\\
        &\quad
                +\ell (\ell +1)
                        \biggl[
                                \left(\pd_r A^{({\rm e})\ell m} -A_r^{\ell m}\right)^2
                                -\left(\pd_\eta A^{({\rm e})\ell m}-A_\eta^{\ell m}\right)^2
                                -a^2m_A^2\left( A^{({\rm e})\ell m}\right)^2
                        \biggr]
                        \biggr\}
        \,,\label{eq:even mode action}
}
is the action for the even parity modes, and
\al{
        &S^{({\rm o})}
                =\frac{1}{2}\sum_{\ell m}
                        \ell (\ell +1)\int\dd r\dd\eta
                        \biggl\{
                                \left(\pd_r A^{({\rm o})\ell m}\right)^2
                                -\left(\pd_\eta A^{({\rm o})\ell m}\right)^2
                                -\left(\frac{\ell (\ell +1)}{\cosh^2 r}+a^2m_A^2\right)\left( A^{({\rm o})\ell m}\right)^2
                        \biggr\}
        \,,\label{eq:odd mode action}
}
is the action for the odd parity modes.

Since the U(1) gauge field contains an auxiliary variable, or
non-dynamical degree of freedom, we need to remove it to find the
appropriate KG norm that contains only the physical degrees of freedom. 
In the region-C, $A_r$ rather than $A_\eta$ behaves as the auxiliary variable
which does not have the time kinetic term. Varying the action with
respect to $A_r$\,, we have the constraint equation, which is given by 
\al{
        \hat\mcO A_r^{\ell m}
                \equiv
                \biggl[-\pd_\eta^2 +m_A^2a^2+\frac{\ell (\ell +1)}{\cosh^2 r}\biggr] A_r^{\ell m}
                =\frac{\ell (\ell +1)}{\cosh^2 r}\pd_r A^{({\rm e})\ell m}-\pd_\eta\pd_r A_\eta^{\ell m}
        \,.\label{eq:constraint eq}
}

The even parity modes contain two degrees of freedom. One of them is the
$3$-dimensional scalar mode, and the other is the $3$-dimensional even
parity vector mode. In order to decompose the action properly we
introduce the even parity vector-type variable by 
\al{
        V^{({\rm e})\ell m}
                \equiv A^{({\rm e})\ell m}+\hat\mcK^{-1}\pd_\eta A_\eta^{\ell m}
        \,,\label{eq:V^e def}
}
where $\hat\mcK$ is a derivative operator given by
\al{
        \hat\mcK =-\pd_\eta^2 +m_A^2a^2
        \,.
}
We then switch 
$\{A_\eta^{\ell m} ,A^{({\rm e})\ell m}\}$ to 
$\{A_\eta^{\ell m}, V^{({\rm e})\ell m}\}$
as dynamical degrees of freedom.
In terms of the new set of variables, namely 
$\{A_\eta^{\ell m},V^{({\rm e})\ell m}\}$\,, 
the constraint equation \eqref{eq:constraint eq} can be rewritten as 
\al{
        A_r^{\ell m}
                =\hat\mcO^{-1}\frac{\ell (\ell +1)}{\cosh^2 r}\pd_r V^{({\rm e})\ell m}
                        -\hat\mcK^{-1}\pd_\eta\pd_r A_\eta^{\ell m}
        \,.\label{eq:constraint eq 2}
}
Note that we need to specify a boundary condition to properly define 
$\hat\mcK^{-1}\pd_\eta A_\eta^{\ell m}$ in eq.~\eqref{eq:V^e def} since
it contains an inverse operator. Different boundary conditions would
lead to different prescriptions for the decomposition of 
$A^{({\rm e})\ell m}$. Note that the boundary condition for
$\hat\mcK^{-1}\pd_\eta A_\eta^{\ell m}$ must be consistent with the
boundary condition for the source $\pd_\eta A_\eta^{\ell m}$ but
otherwise can be specified arbitrarily for our convenience. We have
already imposed the boundary condition that 
$A_\eta^{\ell m}\,,A_r^{\ell m}\,, A^{({\rm e})\ell m}$ decay as fast as
(or faster than) $e^{-|\eta|}$ at $\eta\rightarrow\pm\infty$\,. In
particular this boundary condition for $A_\eta^{\ell m}$ makes it
possible for us to impose the boundary condition that 
$\hat\mcK^{-1}\pd_\eta A_\eta^{\ell m}$ also decay as fast as 
(or faster than) $\propto e^{-|\eta |}$ at
$\eta\rightarrow\pm\infty$\,. The boundary condition for 
$A^{({\rm e})\ell m}$ then implies that 
$V^{({\rm e})\ell m}$ also decays as fast as (or faster than) 
$\propto e^{-|\eta |}$ at $\eta\rightarrow\pm\infty$\,.

Substituting eqs.~\eqref{eq:V^e def} and \eqref{eq:constraint eq 2} into
eq.~\eqref{eq:even mode action}\,, after lengthly calculation, we obtain
the reduced action only for the dynamical degrees of freedom. The
resultant reduced action is given by 
\al{
        S^{({\rm e})}=S^{({\rm e})}_{\rm s}+S^{({\rm e})}_{\rm v}
        \,,\label{eq:reduced action}
}
where
\al{
        S^{({\rm e})}_{\rm s}
                =&\frac{1}{2}\sum_{\ell m}\int\dd r\dd\eta
                        \biggl\{
                                \cosh^2 r\left(\pd_r A_\eta^{\ell m}\right)
                                \Bigl[ 1+\pd_\eta\hat\mcK^{-1}\pd_\eta\Bigr]
                                \left(\pd_r A_\eta^{\ell m}\right)
        \notag\\
        &\quad\quad
                                -\ell (\ell +1)A_\eta^{\ell m}\Bigl[ 1+\pd_\eta\hat\mcK^{-1}\pd_\eta\Bigr] A_\eta^{\ell m}
                                -m_A^2a^2\cosh^2 r\left( A_\eta^{\ell m}\right)^2
                        \biggr\}
        \,,
}
for the scalar mode, and
\al{
        S^{({\rm e})}_{\rm v}
                =&\frac{1}{2}\sum_{\ell m}\ell (\ell +1)\int\dd r\dd\eta
                        \biggl\{
                                \left(\pd_r V^{({\rm e})\ell m}\right)
                                \biggl[ 1-\hat\mcO^{-1}\frac{\ell (\ell +1)}{\cosh^2 r}\biggr]
                                \left(\pd_r V^{({\rm e})\ell m}\right)
                                -V^{({\rm e})\ell m}\hat\mcK\, V^{({\rm e})\ell m}
                        \biggr\}
        \,,
}
for the vector mode. Here, we have used the boundary conditions for 
$A_\eta^{\ell m}$ and $V^{({\rm e})\ell m}$ to show that some boundary
terms, such as those including the interaction between $A_\eta^{\ell m}$ 
and $V^{({\rm e})\ell m}$, vanish. Hence the scalar and vector modes are
completely decoupled in the action for the even parity mode.

We can now define the KG norm by using the reduced actions obtained
above, following and extending \cite{Mukohyama:1999kj}.~\footnote{As we shall see below, an equivalent method to define
the KG norm is provided by a general formula derived in Appendix 
\ref{sec:KG norm}. An advantage of this alternative method is that it
can be applied without eliminating auxiliary fields in the action. The
result is of course the same, as far as the boundary conditions
specified above are imposed.}
To quantize the system of the U(1) gauge
field, we promote  the physical degrees of freedom ${\bm A}$ to
operators $\hat{\bm A}$\,, and expand $\hat{\bm A}$ by mode functions
$\{{\bm A}_\mcN\,,\overline{{\bm A}_\mcN}\}$ as 
\al{
        \hat{\bm A}(x)
                =\sum_\mcN\biggl[\hat a_\mcN{\bm A}_\mcN (x)+\hat a_\mcN^\dagger\overline{{\bm A}_\mcN (x)}\biggr]
        \,,
}
where $\hat a_\mcN$ and $\hat a_\mcN^\dagger$ are the annihilation and
creation operators, respectively, that satisfy the commutation relation,
$[\hat a_\mcN,\hat a_\mcM^\dagger ]=\delta_{\mcN\mcM}$\,. The 
quantum fluctuations of the field are described by the vacuum state
$|0\rangle$ such that $\hat a_\mcN|0\rangle =0$ for any $\mcN$\,. We
note that $\{{\bm A}_\mcN\,,\overline{{\bm A}_\mcN}\}$ should 
form a complete set of linearly independent solutions of the equation of
motion. With these variables, we define the KG norms as 
\al{
        &\left({\bm A}_\mcN ,{\bm A}_\mcM\right)_{\rm KG}^{({\rm s})}
        \notag\\
        &
                =-i\cosh^2 r\int\dd\eta\dd\Omega
                        \biggl\{
                                A_{\eta ,\mcN}\Bigl[ 1+\pd_\eta\hat\mcK^{-1}\pd_\eta\Bigr]\pd_r\overline{A_{\eta ,\mcM}}
                                -\Bigl[ 1+\pd_\eta\hat\mcK^{-1}\pd_\eta\Bigr]\pd_r A_{\eta ,\mcN}\overline{A_{\eta ,\mcM}}
                        \biggr\}
        \,,\label{eq:even scalar KG norm}
}
for the even parity scalar modes,
\al{
        &\left({\bm A}_\mcN ,{\bm A}_\mcM\right)_{\rm KG}^{({\rm v})}
        \notag\\
        &
                =-i\ell (\ell +1)\int\dd\eta\dd\Omega
                        \biggl\{
                                V^{({\rm e})}_\mcN\biggl[ 1-\hat\mcO^{-1}\frac{\ell (\ell +1)}{\cosh^2 r}\biggr]\pd_r\overline{ V^{({\rm e})}_\mcM}
                                -\biggl[ 1-\hat\mcO^{-1}\frac{\ell (\ell +1)}{\cosh^2 r}\biggr]\pd_r V^{({\rm e})}_\mcN\overline{V^{({\rm e})}_\mcM}
                        \biggr\}
        \,,\label{eq:even vector KG norm}
}
for the even parity vector modes, and
\al{
        &\left({\bm A}_\mcN,{\bm A}_\mcM\right)_{\rm KG}^{({\rm o})}
                =-i\ell (\ell +1)\int\dd\eta\dd\Omega
                        \biggl\{
                                A_\mcN^{({\rm o})}\pd_r\overline{A_\mcM^{({\rm o})}}
                                -\left(\pd_r A_\mcN^{({\rm o})}\right)\overline{A_\mcM^{({\rm o})}}
                        \biggr\}
        \,,\label{eq:odd KG norm}
}
for the odd parity modes.
With the KG norm defined above, all modes should be properly normalized
on a Cauchy surface as 
\al{
        \left({\bm A}_\mcN\,,{\bm A}_\mcM\right)_{\rm KG}^{(\lambda )}
                =\delta_{\mcN\mcM}
        \,,\label{eq:normalization}
}
with $\lambda ={\rm s}\,,{\rm v}\,,{\rm o}$\,.
Once $A_\eta^{\ell m}$, $V^{({\rm e})\ell m}$ and $A_\eta^{(o)}$ 
are properly evaluated by solving the equation of motion, we can
calculate the KG norm through 
eqs.~\eqref{eq:even scalar KG norm}-\eqref{eq:normalization}.

In some cases it is convenient to rewrite the reduced action and the KG
norm in terms of auxiliary fields. Introducing the new auxiliary fields, 
$\mcS_r^{\ell m}$ and $V_r^{\ell m}$\,, which obey the constraint
equations: 
\al{
        \hat\mcK\,\mcS_r^{\ell m}=-\pd_\eta\pd_r A_\eta^{\ell m}
        \,,\ \ \ 
        \hat\mcO\, V_r^{\ell m}=\frac{\ell (\ell +1)}{\cosh^2 r}\pd_r V^{({\rm e})\ell m}
        \,,\label{eq:constraint eq 3}
}
$A_r^{\ell m}$ in eq.~\eqref{eq:constraint eq 2} can be reduced to
\al{
        A_r^{\ell m}=V_r^{\ell m}+\mcS_r^{\ell m}
        \,,
}
and we can use $\mcS_r^{\ell m}$ and $V_r^{\ell m}$
as the auxiliary fields for the scalar and vector modes rather than
$A_r^{\ell m}$. With these variables, the reduced actions for the
scalar- and vector-modes are rewritten as
\al{
        S^{({\rm e})}_{\rm sca}
                =&\frac{1}{2}\sum_{\ell m}\int\dd r\dd\eta
                        \biggl\{
                                \cosh^2 r\left(\pd_r A_\eta^{\ell m}-\pd_\eta\mcS_r^{\ell m}\right)^2
                                +m_A^2a^2\cosh^2 r\left(\mcS_r^{\ell m}\right)^2
        \notag\\
        &\quad\quad
                                -\ell (\ell +1)A_\eta^{\ell m}\Bigl[ 1+\pd_\eta\hat\mcK^{-1}\pd_\eta\Bigr] A_\eta^{\ell m}
                                -m_A^2a^2\cosh^2 r\left( A_\eta^{\ell m}\right)^2
                        \biggr\}
        \,,\\
        S^{({\rm e})}_{\rm vec}
                =&\frac{1}{2}\sum_{\ell m}\int\dd r\dd\eta
                        \biggl\{
                                \ell (\ell +1)\left(\pd_r V^{({\rm e})\ell m}-V_r^{\ell m}\right)^2
        \notag\\
        &\quad\quad
                                +\cosh^2 rV_r^{\ell m}\hat\mcK V_r^{\ell m}
                                -\ell (\ell +1)V^{({\rm e})\ell m}\hat\mcK V^{({\rm e})\ell m}
                        \biggr\}
        \,.
}
Following the same step as discussed in Appendix \ref{sec:KG norm}, we
can define the KG norm in terms of the auxiliary fields as 
\al{
        &\left({\bm A}_\mcN ,{\bm A}_\mcM\right)_{\rm KG}^{({\rm s})}
                =-i\cosh^2 r\int\dd\eta\dd\Omega
                        \biggl\{
                                A_{\eta ,\mcN}\left(\pd_r\overline{A_{\eta ,\mcM}}-\pd_\eta\overline{\mcS_{r,\mcM}}\right)
                                -\left(\pd_r A_{\eta ,\mcN}-\pd_\eta\mcS_{r,\mcN}\right)\overline{A_{\eta ,\mcM}}
                        \biggr\}
        \,,\\
        &\left({\bm A}_\mcN ,{\bm A}_\mcM\right)_{\rm KG}^{({\rm v})}
                =-i\ell (\ell +1)\int\dd\eta\dd\Omega
                        \biggl\{
                                V^{({\rm e})}_\mcN\left(\pd_r\overline{ V^{({\rm e})}_\mcM}-\overline{V_{r,\mcM}}\right)
                                -\left(\pd_r V^{({\rm e})}_\mcN -V_{r,\mcN}\right)\overline{V^{({\rm e})}_\mcM}
                        \biggr\}
        \,,
}
for the even parity scalar and vector modes, respectively, 
where the auxiliary fields $\mcS_r^{\ell m}$ and $V_r^{\ell m}$ 
are determined by the constraint equation \eqref{eq:constraint eq 3}. 
It is easy to see that these expressions for the KG norm are equivalent
to \eqref{eq:even scalar KG norm}-\eqref{eq:normalization}.

\section{Mode functions}
\label{sec: Mode functions}

In this section we construct a complete set of mode functions. Since odd
and even parity sectors are decoupled, we shall investigate each sector
separately.

\subsection{Odd parity modes}

First we consider the odd parity sector. The odd parity sector contains
one dynamical degree of freedom, which corresponds to the odd parity
part of a $3$-dimensional transverse vector. Let us construct a set of
positive frequency functions corresponding to the variable 
$A^{({\rm o})}$\,. Varying the action \eqref{eq:odd mode action} with
respect to $A^{({\rm o})}$\,, we obtain
\al{
        \biggl[
                \pd_r^2 -\pd_\eta^2 +\frac{\ell (\ell +1)}{\cosh^2 r}+m_A^2a^2
        \biggr] A^{({\rm o})\ell m}=0
}
In order to solve this equation, we expand $A^{({\rm o})}$ as
\al{
        A^{({\rm o})\ell m}(\eta ,r)
                =\sum_p v^{({\rm o})}_p(\eta )\left(\frac{1}{\sqrt{\ell (\ell +1)}}\cosh rf^{p\ell}(r)\right)
        \,,
}
where the ``summation'' on the r.h.s. should be understood as the
integral over continuum modes ($p^2>0$) plus the summation over discrete
modes ($p^2<0$), if any. Here, we have fixed the coefficient in front of
$f^{p\ell}(r)$ so that the expression inside the parenthesis, when
multiplied by $Y_{\ell m:b}\epsilon^b{}_a$\,, corresponds to
the odd-mode vector-type harmonic function on a unit $3$-hyperboloid
analytically-continued to the region-C (see eqs.~\eqref{eq:Odd
normalization} and \eqref{eq:r relation}). Appendix \ref{sec:Harmonics
in open universe} summarizes the characteristics of the scalar- and
vector-type harmonic functions in the open universe. The equation for
$f^{p\ell}$ is given by 
\al{
        \biggl[
                -\frac{1}{\cosh^2 r}\frac{\dd}{\dd r}\left(\cosh^2 r\frac{\dd}{\dd r}\right)
                -\frac{\ell (\ell +1)}{\cosh^2 r}
        \biggr] f^{p\ell}(r)=(p^2+1)f^{p\ell}(r)
        \,.\label{eq:f eq}
}
Adopting the Euclidean vacuum state as a natural choice after quantum
tunneling, we impose that the positive frequency functions are regular
at $r=0$\, (see eq.~\eqref{eq:f_pl 1}). We then have the explicit
expression for $f^{p\ell}$ as 
\al{
        f^{p\ell}(r)
                =\sqrt{\frac{\Gamma (ip+\ell +1)\Gamma (-ip+\ell +1)}{i\Gamma (ip)\Gamma (-ip)\cosh r}}
                        P^{-\ell -\frac{1}{2}}_{ip-\frac{1}{2}}(i\sinh r)
        \,,\label{eq:f^pl}
}
where $P^\mu_\nu$ is the associated Legendre function of the first kind, 
and we have fixed the normalization constant so that the analytic continuation
of $Y^{p\ell m}(r,\Omega )\equiv f^{p\ell}(r)Y_{\ell m}(\Omega )$
to the region-R or L behaves as a harmonic function properly normalized
on a unit $3$-hyperboloid 
(see Appendix \ref{sec:Harmonics in open universe}).

In this expression, $v^{({\rm o})}$ is an eigenfunction of the operator
$\hat\mcK$ with the eigenvalue $p^2$, that is, $v^{({\rm o})}_p$
satisfies 
\al{
        \hat\mcK v^{({\rm o})}_p
                =\biggl[-\frac{\dd^2}{\dd\eta^2}+m_A^2a^2\biggr] v^{({\rm o})}_p=p^2v^{({\rm o})}_p
        \,.\label{eq:v^o eq}
}
The boundary condition for $A^{({\rm o})\ell m}$ is that 
$v^{({\rm o})}_p$ should decay as fast as (or faster than) 
$e^{-|\eta |}$ at $\eta\rightarrow \pm\infty$. Since the effective
potential $m_A^2a^2$ is clearly positive definite, this in particular
implies that there is no solution with negative $p^2$. It is thus
concluded that there is no supercurvature mode ($p^2<0$ mode) in the odd 
parity sector.

To find the two independent solutions for $v^{({\rm o})}_p$ with
$p^2> 0$, it is useful to introduce the two normalized orthogonal
solutions $\varpi_{\pm ,p}$ which satisfy 
\al{
        \hat\mcK\varpi_{\pm ,p}
                =\biggl[-\frac{\dd^2}{\dd\eta^2}+m_A^2a^2\biggr]\varpi_{\pm ,p}=p^2\varpi_{\pm ,p}
        \,.\label{eq:varpi eq}
}
Since $a=1/H\cosh\eta$\,, it is easy to solve this equation, and the
general solution is 
\al{
        \varpi_{\pm ,p}
                =C_{1,p}^{\pm}\,P_{\nu'}^{ip}(-\tanh\eta )
                        +C_{2,p}^{\pm}\,P_{\nu'}^{-ip}(-\tanh\eta )
        \,,\label{eq:exact sol}
}
where $C_{1,p}^{\pm}$ and $C_{2,p}^{\pm}$ are constants, 
\al{
        &\nu' =\sqrt{\frac{9}{4}-\frac{M_{\rm eff}^2}{H^2}}-\frac{1}{2}
        \,,\ \ 
        M_{\rm eff}^2=m^2+2H^2
        \,. \label{eq:nu'}
}
To construct the independent solutions, let us consider the scattering
problem for $\varpi_{\pm ,p}$. Since the solutions asymptotically
approach linear combinations of the plane waves $e^{\pm ip\eta}$ as
$\eta\to\pm\infty$\,, the equation \eqref{eq:varpi eq} describes
incident plane waves interacting with the effective potential $m_A^2a^2$
and producing reflected and transmitted waves to $\eta\to\pm\infty$ and
$\eta\to\mp\infty$\,, respectively. We then take the two independent
solutions having the following asymptotic behaviors: 
\al{
        \varpi_{+,p}
                \rightarrow \Biggl\{
                        \begin{array}{ll}
                                \rho_{+,p}e^{+ip\eta}+e^{-ip\eta} &:\  \eta\rightarrow +\infty\\
                                \sigma_{+,p}e^{-ip\eta} &:\  \eta\rightarrow -\infty\\
                        \end{array}
        \,,\label{eq:varpi+}\\
        \varpi_{-,p}
                \rightarrow \Biggl\{
                        \begin{array}{ll}
                                \rho_{-,p} e^{-ip\eta}+e^{+ip\eta} &:\  \eta\rightarrow -\infty\\
                                \sigma_{-,p}e^{+ip\eta} &:\  \eta\rightarrow +\infty\\
                        \end{array}
        \,.\label{eq:varpi-}
}
The reflection and transmission coefficients satisfy the following
Wronskian relations~\cite{Garriga:1998he}: 
\al{
        |\rho_{\pm ,p}|^2+|\sigma_{\pm ,p}|^2=1
        \,,\ \ 
        \sigma_{+,p}=\sigma_{-,p}
        \,,\ \ 
        \sigma_{+,p}\overline{\rho_{-,p}}+\overline{\sigma_{-,p}}\rho_{+,p}=0
        \,.\label{eq:Wronskian relations}
}
These solutions are shown to be orthogonal,  
\al{
        \int^\infty_{-\infty}\dd\eta w_{\sigma ,p}\overline{w_{\sigma' ,p'}}
                =2\pi\delta_{\sigma\sigma'}\delta_{\rm D} (p-p')
        \,.\label{eq:orthognality condition}
}
Comparing the asymptotic behavior of the exact solution \eqref{eq:exact sol}
and eqs.~\eqref{eq:varpi+}-\eqref{eq:Wronskian relations}\,,
we find the corresponding coefficients as
\al{
        &C_{1,p}^+
                =\frac{\Gamma (-ip-\nu' )\Gamma (1-ip+\nu' )}{\Gamma (-ip)}
        \,,\ \ 
        C_{2,p}^+
                =0
        \,, \label{eq:C_2,p^+}\\
        &C_{1,p}^-
                =\frac{\sin (\pi\nu' )}{\pi}\Gamma (1-ip)\Gamma (-ip-\nu' )\Gamma (1-ip+\nu' )
        \,,\ \ 
        C_{2,p}^-
                =\Gamma (1-ip)
        \,.\label{eq:C_2,p^-}
}
The two independent solutions for the odd parity modes are expressed in
terms of these solutions, namely 
$v^{({\rm o})}_{\pm,p}\propto\varpi_{\pm ,p}$\,. If $\eta$ were the time
variable then either one of $e^{\pm ip\eta}$ would be chosen by a
boundary condition to specify a quantum state of the system (e.g. the  
Bunch-Davies vacuum). However, in the present situation, it is $r$ that
is the time variable and thus a quantum state of the system is chosen by
imposing a boundary condition on the function of $r$ as in
eq.~\eqref{eq:f^pl}\,. Hence both of $e^{\pm ip\eta}$ should be treated
as independent mode functions and are needed for the construction of a
complete set of mode functions. In order to quantize the perturbations, 
we introduce a variable defined by 
\al{
        A^{({\rm o})}_{\sigma p\ell m}(\eta ,r,\Omega )
                =N_p^{({\rm o})}\varpi_{\sigma ,p}(\eta )\left(\frac{1}{\sqrt{\ell (\ell +1)}}\cosh rf^{p\ell}(r)Y_{\ell m}(\Omega )\right)
        \,, \label{eq:odd-modefunction}
}
where $N_p^{({\rm o})}$ is a normalization constant. Recalling that we
have already proven the absence of supercurvature modes in the odd parity 
sector (see eq.~\eqref{eq:v^o eq}) and substituting this into
eq.~\eqref{eq:odd KG norm}\,, we require 
\al{
        ({\bm A}_{\sigma p\ell m},{\bm A}_{\sigma' p'\ell' m'})^{({\rm o})}_{\rm KG}
                =&4p\sinh (\pi p)\left( N^{({\rm o})}_p\right)^2\delta_{\sigma\sigma'}\delta_{\rm D}(p-p')\delta_{\ell\ell'}\delta_{mm'}
        \notag\\
                =&\delta_{\sigma\sigma'}\delta_{\rm D} (p-p')\delta_{\ell\ell'}\delta_{mm'}
        \,,
}
where we have used the orthogonality condition for $\varpi_{\sigma ,p}$
(see eq.~\eqref{eq:orthognality condition})\,. Hence the KG
normalization condition implies that the normalization constant is given
by 
\al{
        N^{({\rm o})}_p=\frac{1}{2\sqrt{p\sinh (\pi p)}}
        \,. \label{eq:odd-normalization}
}

In summary, we have shown that there is no supercurvature mode in the
odd parity sector and that continuous odd parity modes (with $p^2>0$) 
are given by \eqref{eq:odd-modefunction} with
\eqref{eq:odd-normalization}, \eqref{eq:exact sol}-\eqref{eq:nu'} and
\eqref{eq:C_2,p^+}-\eqref{eq:C_2,p^-}.

\subsection{Even parity modes}

In this subsection, we construct a complete set of mode functions in the
even parity sector. The equations for $A_\eta^{\ell m}$ and 
$V^{({\rm e})\ell m}$ are given by 
\al{
        &\biggl[-\frac{1}{\cosh^2 r}\pd_r\left(\cosh^2 r\pd_r\right) -\frac{\ell (\ell +1)}{\cosh^2 r}\biggr]
                \left( 1+\pd_\eta\hat\mcK^{-1}\pd_\eta\right) A_\eta^{\ell m}
                =m_A^2a^2A_\eta^{\ell m}
        \,,
}
for the $3$-dimensional scalar modes, and
\al{
        \hat\mcK\, V^{({\rm e})\ell m}
                =-\pd_r\biggl[\left( 1-\hat\mcO^{-1}\frac{\ell (\ell +1)}{\cosh^2 r}\right)\pd_r V^{({\rm e})\ell m}\biggr]
        \,,
}
for the $3$-dimensional vector modes. Assuming $m_A^2\neq 0$\,, we then
expand $A_\eta^{\ell m}$ and $V^{({\rm e})\ell m}$ as 
\al{
        &A_\eta^{\ell m}(\eta ,r)
                =\sum_p\frac{\chi_p (\eta )}{a(\eta )}f^{p\ell}(r)
        \,,\\
        &V^{({\rm e})\ell m}(\eta ,r)
                =\sum_p v_p^{({\rm e})}(\eta )
                        \biggl[
                                \frac{1}{\sqrt{\ell (\ell +1)}p}
                                \frac{\dd}{\dd r}\left( \cosh r\,f^{p\ell}(r)\right)
                        \biggr]
        \,,
}
where the coefficients have been fixed so that the analytic
continuations of these functions to region-R or L corresponds to the
scalar- and vector-type harmonic functions defined in Appendix
\ref{sec:Harmonics in open universe}. Thus, the equations for $\chi_p$
and $v^{({\rm e})}_p$ can be reduced to 
\al{
        &\biggl[-\frac{\dd^2}{\dd\eta^2}+\left( m_A^2a^2+a\frac{\dd^2}{\dd\eta^2}\left(\frac{1}{a}\right) -1\right)\biggr]\chi_p =
        \biggl[-\frac{\dd^2}{\dd\eta^2}+m_A^2a^2\biggr]\chi_p 
                =p^2\chi_p
        \,,\\
        &\biggl[-\frac{\dd^2}{\dd\eta^2}+m_A^2a^2\biggr] v_p^{({\rm e})}=p^2v_p^{({\rm e})}
        \,.
}
These equations for $\chi_p$ and $v^{({\rm e})}_p$ are exactly the same
as that for $\varpi_{\pm ,p}$ investigated in the previous subsection
(see eqs.~\eqref{eq:varpi eq}-\eqref{eq:C_2,p^-}). The boundary
conditions for $A_\eta^{\ell m}$ and $V^{({\rm e})\ell m}$ imply that
$\chi_p$ and $v_p^{({\rm e})}$ decay as fast as (or faster than)
$e^{-2|\eta |}$ and $e^{-|\eta |}$, respectively, at 
$\eta\rightarrow \pm\infty$. It is thus concluded that there is no
supercurvature mode in the even parity sector for the same reason as in
the odd parity sector, i.e. because of the positivity of the effective
potential $m_A^2a^2$\,.

It is also straightforward to repeat the same procedure as in the
previous subsection to find a complete set of continuous ($p^2>0$) mode
functions in the even parity sector. Mode functions are simply expressed
in terms of $\varpi_{\pm ,p}$\,. To quantize the perturbations, we
introduce variables: 
\al{
        &A_{\eta ,\sigma p\ell m}(\eta ,r,\Omega )
                =N^{({\rm s})}_p\frac{\varpi_{\sigma ,p} (\eta )}{a(\eta )}f^{p\ell}(r)Y_{\ell m}(\Omega )
        \,,\label{eq:A_eta variable}\\
        &V^{({\rm e})}_{\sigma p\ell m}(\eta ,r,\Omega )
                =N_p^{({\rm v})}\varpi_{\sigma ,p}(\eta )
                        \biggl[
                                \frac{1}{\sqrt{\ell (\ell +1)}p}
                                \frac{\dd}{\dd r}\left( \cosh r\,f^{p\ell}(r)\right) Y_{\ell m}(\Omega )
                        \biggr]
        \,,\label{eq:V^e variable}
}
where $N^{({\rm s})}_p$ and $N^{({\rm e})}_p$ are normalization
constants to be determined for the scalar and vector modes,
respectively. We can determine the normalization constants by using the
KG norms. Substituting eqs.~\eqref{eq:A_eta variable} and 
\eqref{eq:V^e variable} into the KG norm for the even parity modes
defined in eqs.~\eqref{eq:even scalar KG norm} and 
\eqref{eq:even vector KG norm}, we have
\al{
        \left({\bm A}_{\sigma p\ell m},{\bm A}_{\sigma' p'\ell' m'}\right)^{({\rm s})}_{\rm KG}
                =&m_A^2\frac{4p\sinh (\pi p)}{p^2+1}\left( N_p^{({\rm s})}\right)^2
                        \delta_{\sigma\sigma'}\delta_{\rm D}(p-p')\delta_{\ell\ell'}\delta_{mm'}
        \,,
}
for the $3$-dimensional scalar mode, and 
\al{
        \left({\bm A}_{\sigma p\ell m},{\bm A}_{\sigma' p'\ell' m'}\right)^{({\rm v})}_{\rm KG}
                =&4p\sinh (\pi p)\left( N_p^{({\rm v})}\right)^2
                        \delta_{\sigma\sigma'}\delta_{\rm D}(p-p')\delta_{\ell\ell'}\delta_{mm'}
        \,,
}
for the $3$-dimensional vector mode. When we require the normalization
condition eq.~\eqref{eq:normalization}, the scalar- and vector-modes are
normalized respectively as 
\al{
        &N_p^{({\rm s})}=\frac{1}{2m_A}\sqrt{\frac{p^2+1}{p\sinh (\pi p)}}
        \,,\ \ \ 
        N_p^{({\rm v})}=\frac{1}{2\sqrt{p\sinh (\pi p)}}
        \,.\label{eq:even parity mode normalization conditon}
}

In summary, we have found that there is no supercurvature mode in 
the even parity sector and that a complete set of even parity continuous
mode functions is given by \eqref{eq:A_eta variable}, \eqref{eq:V^e variable} with 
\eqref{eq:even parity mode normalization conditon}.

\section{Consistency of neutral case and decoupling limit}
\label{sec: massless limit}

In the previous section, we have shown that there is no supercurvature
mode for a U(1) gauge field with both gauge and conformal symmetries 
spontaneously broken through the Higgs mechanism, for any values of the
mass of the vector field.

It has been known that a scalar field $\phi$ with a sufficiently light
effective mass, $0\leq m<\sqrt{2}H$, has a supercurvature
mode~\cite{Sasaki:1994yt}, and the supercurvature mode survives the
massless limit. Furthermore, the existence of the supercurvature mode is
essential for the recovery of the correct massless limit of the Wightman
function. On one hand, one can show that the Euclidean Wightman function
in the limit $m\to 0$ contains a constant divergent
contribution~\cite{Sasaki:1994yt}: 
\al{
        \lim_{m\rightarrow 0}\ave{0|\phi (x)\phi (x')|0}=\frac{3H^4}{8\pi^2 m^2} + \mathcal{O}(m^0)
        \,.\label{scalar diverge}
}
On the other hand, the contribution of all subcurvature modes (with
$p^2>0$) to the Wightman function remains finite in the massless
limit. This means that the set of all subcurvature modes does not form a
complete set of mode functions and that something is missing. It is the
supercurvature mode that is missing here. The contribution from the
supercurvature mode (with $p^2=-1$) correctly reproduces the divergent 
behavior of the Euclidean Wightman function shown in 
eq.~\eqref{scalar diverge}.

From the above observation on the Wightman function of a scalar field, it
is expected that the massless limit serves as a useful consistency
check also for vector fields. We thus consider the massless limit of the
massive vector field and see whether the correct behavior of the
Wightman function can be reproduced by the contributions from
subcurvature modes only, without need for any supercurvature modes.

For the system of the U(1) gauge field considered in the present paper,
the massless limit is provided by the decoupling limit, i.e. the 
$e\to 0$ limit. As we shall see in the following, the decoupling limit
appears to be rather confusing. On one hand, we have shown that the
massive vector field does not have a supercurvature mode for any
non-zero value of $e$. On the other hand, for $e=0$, i.e. if the
(would-be) Higgs field is neutral under the U(1), the system consisting
of the U(1) gauge field and the phase of the complex (would-be) Higgs
field is reduced to a massless vector field plus a massless scalar
field. Since a massless scalar field is known to have a supercurvature
mode, there appears discontinuity in the $e\to 0$ limit. We need to
reconcile these two apparently contradicting results.

In this section we first reconcile the apparent contradiction between
the $e\to 0$ limit of the $e\ne 0$ theory and the $e=0$ theory
(subsection \ref{sec:exact massless}). We then investigate the Wightman
function in the decoupling ($e\to 0$) limit as a consistency check
(subsection \ref{sec: massive gauge}).

\subsection{Neutral ($e=0$) case}
\label{sec:exact massless}

Before considering the decoupling ($e\to 0$) limit, let us investigate
the neutral ($e=0$) case. In this subsection, we focus only on the
scalar sector since the apparent contradiction explained above is in
this sector. One can begin with the action 
eq.~\eqref{eq:relevant action} with $e=0$ to derive the equation of
motion of the scalar degree of freedom. Adopting the gauge condition
$A_\eta =0$ for convenience, we obtain the action for the phase
of the (would-be) Higgs field as 
\al{
        &S^{({\rm e})}
        \supset\frac{1}{2}\varphi^2\sum_{\ell m}\int\dd r\dd\eta
                        a^2\cosh^2 r
                \biggl\{
                \left(\pd_r\Theta^{\ell m}\right)^2 -\left(\pd_\eta\Theta^{\ell m}\right)^2
                -\frac{\ell (\ell +1)}{\cosh^2 r}\left( \Theta^{\ell m}\right)^2
                \biggr\}
        \,,
}
where $\Theta^{\ell m}$ is the coefficient of the spherical harmonic
expansion of the phase of the (would-be) Higgs field $\Theta$. Expanding $\Theta^{\ell m}$ in terms of
$f^{p\ell}$ (see eqs.~\eqref{eq:f eq} and \eqref{eq:f^pl}) as
$\Theta^{\ell m}(\eta ,r)=\sum_p\Theta_p (\eta )f^{p\ell}(r)$\,, 
we obtain the equation for $\Theta_p$ as
\al{
        \biggl[\frac{1}{a^2}\frac{\dd}{\dd\eta}\left( a^2\frac{\dd}{\dd\eta}\right) +(p^2+1)\biggr]\Theta_p =0
        \,.
}
Since this is the same as the equation of motion for the massless scalar
field, one might think that there should be a supercurvature mode at
$p^2=-1$\,, according to \cite{Sasaki:1994yt}. However, the solution to
this equation with $p^2=-1$ is trivial, namely 
$\Theta_p=\text{const}$\,, in the entire region-C and turns out to be a
gauge degree of freedom. A key observation here is that in the neutral
($e=0$) case, the U(1) gauge symmetry manifests itself as a global
shift symmetry: $\Theta (x)\rightarrow \Theta (x)+\lambda$, where 
$\lambda$ is a constant. Note that this shift symmetry must be respected
by any interactions including $\Theta$. In particular, observers or 
detectors interacting with $\Theta$ can probe derivatives
$\partial_{\mu}\Theta$ but cannot probe the value of $\Theta$ itself
even in principle. Hence, the constant solution with $p^2=-1$ is not
within the physical spectrum of the theory. In other words, the $p^2=-1$
solution does not affect any correlation functions invariant under the
global shift symmetry since $\Theta$ enters invariant quantities only
through its derivatives.

Therefore it is concluded that there is no supercurvature mode in the
neutral ($e=0$) case. This reconciles the apparent contradiction between
the $e=0$ theory and the $e\to 0$ limit of the $e\ne 0$ theory.

\subsection{Decoupling ($e\to 0$) limit}
\label{sec: massive gauge}

Let us now explore the massless limit of the massive U(1) gauge field,
i.e. the decoupling limit, $e\to 0$, of the theory with $e\ne 0$. In
this subsection we compute the Wightman function of the U(1) gauge field
in the decoupling limit and explicitly verify that the correct behavior
of the Euclidean Wightman function is reproduced by the contributions
from subcurvature modes only, without need for any supercurvature
modes.

The Wightman function for the massive U(1) gauge field in de Sitter
spacetime is previously studied in the 
literature~\cite{Frob:2013qsa,Allen:1985wd, Tsamis:2006gj, Youssef:2010dw}~\footnote{
In \cite{Frob:2013qsa}, the authors found that the Wightman function of
a massive gauge field depends on the way how gauge is fixed. Our gauge
choice corresponds to what they call the Proca theory, and in this gauge
the massless limit of the Wightman function has a simple form. It
seems that the behavior of the Wightman function of a massive gauge
field in de Sitter spacetime is not yet fully
understood~\cite{Allen:1985wd, Tsamis:2006gj, Youssef:2010dw}. 
}.
According to \cite{Frob:2013qsa}, the Wightman function of the massive
U(1) gauge field in the decoupling ($e\to 0$) limit can be written in
terms of the scalar propagator as 
\al{
        \lim_{m_A\rightarrow 0}\ave{0|A_\mu (x)A_{\mu'} (x')|0}
                &=\lim_{m_A\rightarrow 0}\frac{1}{m_A^2}\pd_\mu\pd_{\mu'}\Delta_{M^2} (Z(x,x'))+\mcO (m_A^0)
        \,.\label{eq:gauge Wightman functon}
}
where $Z$ denotes the de Sitter invariant distance between two points,
$x$ and $x'$\,, in de Sitter spacetime, and $\Delta_{M^2}(Z)$ denotes
the propagator of the scalar field with the mass
$M=\sqrt{9/4-(\nu'+3/2)^2}$, which is defined by 
\al{
        \Delta_{M^2}(Z)
                =\frac{H^2}{(4\pi )^2}\frac{\Gamma (3+\nu' )\Gamma (-\nu' )}{\Gamma (2)}
                {}_2F_1\left( 3+\nu' ,-\nu' ;2;\frac{1+Z}{2}\right)
        \,,
}
where ${}_2F_1(a,b;c;z)$ is the hypergeometric function. It should be
noted that the propagator of the massive scalar field in the massless
(decoupling) limit is divergent as seen in eq.~\eqref{scalar diverge}. 
However, the divergent contribution is $Z$-independent and thus drops
out when derivatives are acted on the propagator as in 
\eqref{eq:gauge Wightman functon}. We then take the massless limit and
describe the explicit expression for the divergent contributions of the
Wightman function for the U(1) gauge field as 
\al{
        \lim_{m_A\rightarrow 0}\ave{0|A_\mu (x)A_{\mu'} (x')|0}
                =\frac{H^2}{(4\pi )^2m_A^2}
                        \left[
                                \frac{Z-3}{(Z-1)^3}(\partial_\mu Z)(\partial_{\mu'}Z)-\frac{Z-2}{(Z-1)^2}
                                (\partial_\mu \partial_{\mu'}Z)
                        \right]
        \,.\label{vector diverge}
}
Hereafter we neglect higher-order contributions of order $\mcO (m_A^0)$
for simplicity.

In order to compare the leading-order Wightman function in the
decoupling limit with our result derived in the present paper, we
rewrite eq.~\eqref{vector diverge} in terms of the coordinates in the
open chart of the de Sitter spacetime i.e. the coordinate in the
region-J $(\eta_\rmJ ,r_\rmJ ,\Omega )$\,. The invariant distance $Z$ is
then reduced to 
\al{
     Z(x_\rmJ ,x_\rmJ' )
                =\frac{\cosh \eta_\rmJ \cosh \eta_\rmJ'-\cosh\zeta}{\sinh\eta_\rmJ\sinh\eta_\rmJ'}
        \,,\label{eq:invariant distance in J}
}
where  $\cosh\zeta \equiv \cosh r_\rmJ\cosh r_\rmJ' -\sinh r_\rmJ\sinh r_\rmJ' \cos\Xi$ 
with $\cos\Xi$ being the directional cosine between $\Omega$ and $\Omega'$.
Substituting the invariant distance eq.~\eqref{eq:invariant distance in J}
into eq.~\eqref{vector diverge}, we can easily rewrite each component of
the Wightman function of the U(1) gauge field in term of the coordinate
in the region-J.

We also calculate the Wightman function by using the explicit
expressions for the mode functions derived in section 
\ref{sec: Mode functions}. As an example, let us focus on the 
$(\eta ,\eta' )$-component of the Wightman function of the U(1) gauge 
field. Taking the massless (decoupling) limit ($\nu'\rightarrow 0$) and
the analytic continuation to the region-J, we can rewrite the two
independent solutions for the $\eta$-component of the U(1) gauge field
\eqref{eq:A_eta variable} as 
\al{
        &A_{\eta ,+p\ell m}(\eta_\rmJ ,r_\rmJ ,\Omega )
                =\frac{1}{2m_A}\sqrt{\frac{p^2+1}{p\sinh (\pi p)}}\frac{1}{a_\rmJ (\eta_\rmJ )}\,e^{ip\eta_\rmJ -\pi p/2}f^{p\ell}(r_\rmJ )Y_{\ell m}(\Omega )
        \,,\\
        &A_{\eta ,-p\ell m}(\eta_\rmJ ,r_\rmJ ,\Omega )
                =\frac{1}{2m_A}\sqrt{\frac{p^2+1}{p\sinh (\pi p)}}\frac{\Gamma (1-ip)}{\Gamma (1+ip)}
                        \frac{1}{a_\rmJ (\eta_\rmJ )}\,e^{-ip\eta_\rmJ +\pi p/2}f^{p\ell}(r_\rmJ )Y_{\ell m}(\Omega )
        \,,\label{Apm}
}
where $a_\rmJ (\eta_\rmJ )=-1/H\sinh\eta_\rmJ$\,, and we have used the 
following relation: 
$P_0^{ip}(-\tanh\eta)=e^{ip\eta_R-\pi p/2}/\Gamma (1-ip)$\,.
We can then calculate the $(\eta ,\eta' )$-component of the Wightman
function for the U(1) gauge field as 
\al{
        &\lim_{m_A\rightarrow 0}\ave{0|A_\eta (x)A_{\eta'}(x')|0}
                =\lim_{m_A\rightarrow 0}\sum_{\sigma =\pm}\sum_{p\ell m}
                        A_{\eta ,\sigma p\ell m}(\eta_\rmJ ,r_\rmJ ,\Omega )
                        \overline{A_{\eta ,\sigma p\ell m}(\eta_\rmJ' ,r_\rmJ' ,\Omega' )}
        \notag\\
        &\quad
                =\frac{H^2\sinh\eta_\rmJ\sinh\eta_\rmJ'}{4\pi^2 m_A^2}
                \int_0^\infty\dd p\frac{(p^2+1)\sin (p\zeta )}{\sinh\zeta}
                        \biggl\{
                                \frac{1}{1-e^{-2\pi p}}e^{-ip(\eta_\rmJ -\eta_\rmJ' )}
                                +\frac{e^{-2\pi p}}{1-e^{-2\pi p}}e^{+ip(\eta_\rmJ -\eta_\rmJ' )}
                        \biggr\}
        \notag\\
        &\quad
                =\frac{H^2}{8\pi^2 m_A^2}
                        \sinh\eta_\rmJ\sinh\eta_\rmJ'
                        \frac{2+\cosh^2\zeta -3\cosh\zeta\cosh (\eta_\rmJ -\eta_\rmJ' )}{(\cosh\zeta -\cosh (\eta_\rmJ -\eta_\rmJ' ))^3}
        \,,\label{eq:eta eta Wightman function}
}
where we have used the completeness relation for the scalar harmonics, 
$Y^{p\ell m}(r_\rmJ ,\Omega )\equiv f^{p\ell}(r_\rmJ )Y_{\ell m}(\Omega )$\,,
which is given by
\al{
        &\sum_{\ell m}Y^{p\ell m}(r_\rmJ ,\Omega )\overline{Y^{p\ell m}(r_\rmJ' ,\Omega' )}
                =\frac{p\sin (p\zeta )}{2\pi^2\sinh\zeta}
        \,.
}
One can easily compare the resultant Wightman function 
eq.~\eqref{eq:eta eta Wightman function} with one obtained by
substituting eq.~\eqref{eq:invariant distance in J} into 
\eqref{vector diverge} and find that these leading-order expressions
exactly coincide. This confirms that the $(\eta ,\eta' )$-component of
the Wightman function of the U(1) gauge field in the decoupling limit is
correctly reproduced by the contribution from subcurvature modes only,
without need for supercurvature modes. Following the same step as the  
$(\eta ,\eta')$-component, we can verify the consistency between
eq.~\eqref{vector diverge} and our results for the other components.

\section{Summary} 
\label{sec:summary}

In this paper, we have investigated the Euclidean vacuum mode functions 
of a massive vector field in the spatially open chart of de Sitter
spacetime. In order to clarify whether supercurvature modes exist, we
have studied the U(1) gauge field with gauge and conformal symmetries
spontaneously broken through the Higgs mechanism. We have found that
there is no supercurvature mode for both the even and odd parity
sectors. This implies that it is difficult to generate the sufficient
amount of  the magnetic field on large scales by using the
superadiabatic growth within the one-bubble open inflation scenario even
if the Higgs mechanism spontaneously breaks gauge and conformal
invariances.

Utilizing the obtained mode functions, we have explicitly computed the
Wightman function of the U(1) gauge field in terms of the coordinates in
the open chart of the de Sitter spacetime, and have compared it with one
obtained by other methods. It was found that the leading-order Wightman
function in the decoupling ($e\to 0$) limit is correctly reproduced by
the sum of the products of the subcurvature modes without need for 
introducing supercurvature modes. In consequence we have verified that
the supercurvature mode is not needed as a part of a complete set of
mode functions of the U(1) gauge field in the decoupling limit.

An interesting observation made in subsection \ref{sec:exact massless}
is that the existence/absence of supercurvature modes can be strongly
related to symmetries of the theory. While a massive scalar field with a
sufficiently light mass has a supercurvature mode~\cite{Sasaki:1994yt}
that survives the massless limit, a theory of a scalar field with shift
symmetry does not allow a physical supercurvature mode. This is because
the would-be supercurvature mode does not show up 
in any correlation functions invariant under the shift
symmetry. Furthermore, a vector field with a U(1) gauge symmetry does
not have a supercurvature mode even when the vector field is given a
mass by the Higgs mechanism and thus absorbs a light scalar degree of
freedom (the phase of the complex Higgs field). 
It may be interesting to investigate supercurvature modes of 
the vector field when we take metric perturbations into
account since gravity has the diffeomorphism symmetry, 
although in the present paper we take account of
the effect of gravity only through a curved background. The evaluation
of the metric perturbations is beyond the scope of the present paper and
we hope to come back to this issue in a future publication.

In this paper, we have assumed several simplifications: (i) the universe
during inflationary era after a quantum tunneling is assumed to be well 
approximated by an exact de Sitter spacetime in the open chart; (ii) the
origin of the breaking of the gauge and conformal symmetries and the
mass of the vector field is assumed to be the standard Higgs mechanism;
(iii) the mass squared $V''_{\Phi}(|\Phi|)$ around the minimum of the
Higgs potential is assumed to be large enough so that the mass of the
vector field can be considered as constant during inflation. 
Relaxing some of these assumptions would in principle affect details of
our results, although generic features that we have found are expected
to remain the same. Furthermore, we have neglected the interactions
between the tunneling field and the other fields such as the Higgs
field. If we take into account such interactions, then spatially
localized, bubble-shaped features may appear~\cite{Sugimura:2012kr}. We
hope to come back to these issues in the near future.


\acknowledgments

We would like to thank M.~Sasaki and T.~Tanaka for useful
discussions. D.Y. and T.F. are supported by Grant-in-Aid for JSPS
Fellows Nos.~259800 and 248160. S.M. is supported by Grant-in-Aid for
Scientific Research 24540256 and 21111006. This work was supported by
the World Premier International Research Center Initiative (WPI
Initiative), MEXT, Japan.

\appendix

\section{Intrinsic covariant derivative and reduced action}

In this appendix, we list some useful formulas which we have used in our
calculations. We present the relation between the intrinsic covariant
derivatives on the unit $2$-sphere and the $4$-dimensional covariant
derivatives. We first evaluate the Christoffel symbols of the metric
$\bar g_{\mu\nu}$ defined in eq.~\eqref{eq:metric in region-C}: 
\al{
        &\Gamma^r_{ab}=\cosh r\sinh r\,\omega_{ab}
        \,,\ \ 
        \Gamma^a_{rb}=\tanh r\,\delta^a{}_b
        \,,\\
        &\Gamma^\theta_{\phi\phi}
                =-\sin\theta\cos\theta
        \,,\ \ 
        \Gamma^\phi_{\theta\phi}
                =\cot\theta
        \,,\ \ 
        \text{otherwise}=0
        \,.
}
For the purpose of the $(1+1+2)$ decomposition, we introduce the basis
vectors given by 
\al{
        &u^\mu =\delta^\mu_\eta
        \,,\ \ 
        n^\mu =\delta^\mu_r
        \,,\ \ 
        e^\mu_\theta 
                =\frac{1}{\cosh r}\,\delta^\mu_\theta
        \,,\ \ \ 
        e^\mu_\phi 
                =\frac{1}{\cosh r}\,\delta^\mu_\phi
        \,.\label{eq:basis vectors}
}
We can evaluate the following equations in terms of these basis vectors:
\al{
        &u^\mu\bar\nabla_\mu u^\nu =0
        \,,\ \ \ 
        u^\mu\bar\nabla_\mu n^\nu =0
        \,,\ \ \ 
        u^\mu\bar\nabla_\mu e^\nu_a =0
        \,,\\
        &
        n^\mu\bar\nabla_\mu u^\nu =0
        \,,\ \ \ 
        n^\mu\bar\nabla_\mu n^\nu =0
        \,,\ \ \ 
        n^\mu\bar\nabla_\mu e^\nu_a =0
        \,,\\
        &e^\mu_a\bar\nabla_\mu u^\nu =0
        \,,\ \ \ 
        e^\mu_a\bar\nabla_\mu n^\nu =\tanh r\,e^\nu_a
        \,,\\
        &e^\mu_\phi\bar\nabla_\mu e^\nu_\theta
                =e^\mu_\theta\bar\nabla_\mu e^\nu_\phi =\frac{1}{\cosh r}\cot\theta\, e^\nu_\phi
        \,,\\
        &e^\mu_\theta\bar\nabla_\mu e^\nu_\theta =\tanh r\,n^\nu
        \,,\ \ \ 
        e^\mu_\phi\bar\nabla_\mu e^\nu_\phi 
                =\tanh r\sin^2\theta\, n^\nu
                        -\frac{\sin\theta\cos\theta}{\cosh r}\,e^\nu_\theta
        \,,
}
where $\bar\nabla_\mu$ denotes the covariant derivative with respect to
the conformally related $4$-dimensional metric $\bar g_{\mu\nu}$\,.
With these notations, the $4$-dimensional metric can be decomposed as 
\al{
        \bar g_{\mu\nu}=u_\mu u_\nu -n_\mu n_\nu +\omega_{ab}e^a_\mu e^b_\nu
        \,.
}
We then derive the explicit relation between the covariant derivative of
a two vector on the unit $2$-sphere and the $4$-dimensional covariant derivative as
\al{
        \frac{1}{\cosh r}X_{a:b}
                \equiv &e^\mu_b\bar\nabla_\mu\left( X_\nu e^\nu_a\right)
                                -\left(  e_\nu^c e^\mu_a\bar\nabla_\mu e^\nu_b\right)\left( X_\sigma e^\sigma_c\right)
                =e^\mu_b e^\nu_a\bar\nabla_\mu\left( X_ce_\nu^c \right)
        \,.\label{eq:intrinsic covariant derivative}
}
Using the intrinsic covariant derivative and adopting
the convention to denote the projection of tensors as
\al{
        A_\eta\equiv A_\mu u^\mu
        \,,\ \ \ 
        A_r\equiv A_\mu n^\mu
        \,,\ \ \ 
        \tilde A_a\equiv A_\mu e^\mu_a
        \,,
}
where $a$ runs $\theta$ and $\phi$\,,
the action \eqref{eq:relevant action}
can be rewritten in terms of the $2$-scalars, $A_\eta\,,A_r$\,, and a $2$-vector, $A_a$\,, as
\al{
        S=&\frac{1}{2}\int\dd r\dd\eta\dd\Omega
                \biggl\{
                        \cosh^2 r\left(\pd_r A_\eta -\pd_\eta A_r\right)^2
                        +\omega^{ab}
                                \Bigl[\pd_r\left(\cosh r\,\tilde A_a\right) -A_{r:a}\Bigr]\Bigl[\pd_r\left(\cosh r\,\tilde A_b\right) -A_{r:b}\Bigr]
        \notag\\
        &
                        -\omega^{ab}\left( \cosh r\pd_\eta\tilde A_a-A_{\eta :a}\right)\left( \cosh r\pd_\eta\tilde A_b-A_{\eta :b}\right)
                        -\frac{1}{2}\omega^{am}\omega^{bn}\left(\tilde A_{a:b}-\tilde A_{b:a}\right)\left(\tilde A_{m:n}-\tilde A_{n:m}\right)
        \notag\\
        &
                        +m_A^2a^2\cosh^2 r\left( A_r^2-A_\eta^2 -\omega^{ab}\tilde A_a\tilde A_b\right)
                \biggr\}
        \,.
}
Decomposing the $2$-vector into the even and odd parity modes (see eq.~\eqref{eq:even odd decomposition}), and 
expand the quantities in terms of the spherical harmonics $Y_{\ell m}(\Omega )$ 
(see eqs.~\eqref{eq:spherical expansion 1}, \eqref{eq:spherical expansion 2})\,,
we then obtain the reduced action \eqref{eq:reduced action}-\eqref{eq:odd mode action}
in the main text.

\section{Harmonics in open universe}
\label{sec:Harmonics in open universe}

We briefly summarize the formulas for the scalar and vector harmonics on
the open universe. To characterize the harmonics in open universe, we
introduce the metric on open chart (called region-$\rmJ$ 
($\rmJ =\rmR\,,\rmL$) hereafter), which is defined by
\al{
        \dd s_\rmJ^2
                =a_\rmJ^2 (\eta_\rmJ )
                        \Bigl[
                                -\dd\eta_\rmJ^2 +\gamma_{ij}\dd x^i\dd x^j
                        \Bigr]
                =a_\rmJ^2 (\eta_\rmJ )
                        \Bigl[
                                -\dd\eta_\rmJ^2 +\dd r_\rmJ^2 +\sinh^2 r_\rmJ\omega_{ab}\dd\theta^a\dd\theta^b
                        \Bigr]
        \,,\label{eq:J coordinate}
}
where $a_\rmJ =-1/H\sinh\eta_\rmJ$\,.

\subsection{Scalar harmonics}
\label{sec:Scalar harmonics}

The normalized scalar harmonics, $Y^{p\ell m}$\,, are the eigen function for
the Laplacian operator $\bar\nabla^2$ on the 3-dimensional hyperboloid
in the region-J: 
\al{
        \bar\nabla^2 Y^{p\ell m}+(p^2+1)Y^{p\ell m}=0
        \,,\label{eq:eigen function for Delta}
}
where $\bar\nabla_i$ denotes the covariant derivative with respect to
the three-dimensional metric $\gamma_{ij}$\,, $p$ is the wave number,
$\ell$ and $m$ denote the angular momentum. The scalar harmonics are
expressed in the form 
\al{
        Y^{p\ell m}(r_\rmJ ,\Omega )=f^{p\ell}(r_\rmJ )Y_{\ell m}(\Omega )
        \,,
}
where $Y_{\ell m}(\Omega )$ is the spherical harmonic function on the
unit two-sphere. The equation for $f^{p\ell}$ is given from
eq.~\eqref{eq:eigen function for Delta}: 
\al{
        \Biggl[
                -\frac{1}{\sinh^2 r_\rmJ}\frac{\dd}{\dd r_\rmJ}
                        \left(\sinh^2 r_\rmJ\frac{\dd}{\dd r_\rmJ}\right)
                +\frac{\ell (\ell +1)}{\sinh^2 r_\rmJ}
        \Biggr] f^{p\ell}(r_\rmJ )
        =(p^2+1)f^{p\ell}(r_\rmJ )
        \,.\label{eq:f^pl eq}
}
Requiring the regularity at $r_\rmJ =0$\,, the eigenfunction is given by 
\al{
        f^{p\ell}(r_\rmJ )
                \propto\frac{1}{\sqrt{\sinh r_\rmJ}}
                        P^{-\ell -\frac{1}{2}}_{ip-\frac{1}{2}}(\cosh r_\rmJ )
                                \equiv\mcP_{p\ell}(r_\rmJ )
        \,,\label{eq:f_pl 1}
}
where $P_\nu^\mu$ is the associated Legendre function of the first kind.
For the continuous mode ($p^2>0$)\,, we fix the normalization factor 
so that $Y^{p\ell m}$ satisfies
\al{
        \int\dd r_\rmJ\dd\Omega\,\sinh^2 r_\rmJ\,
                Y^{p\ell m}(r_\rmJ ,\Omega )
                \overline{Y^{p'\ell' m'}(r_\rmJ ,\Omega )}
                =\delta (p-p')\delta_{\ell\ell'}\delta_{mm'}
        \,.\label{eq:normalization condition}
}
Because the divergent contribution at $p=p'$ comes only from
the boundaries of integration at $r_\rmJ =\pm\infty$\,,
the integration can be evaluated without investigating the detailed 
behavior of eq.~\eqref{eq:f_pl 1}\,.
Using the asymptotic behavior of eq.~\eqref{eq:f_pl 1} near the boundaries,
we have
\al{
        &\int^\infty_0\dd r_\rmJ\,\sinh^2 r_\rmJ\,\mcP_{p\ell}(r_\rmJ )\overline{\mcP_{p'\ell}(r_\rmJ )}
        \notag\\
        &\quad\quad
                =\lim_{\epsilon\rightarrow 0}
                        \Biggl[
                                \frac{\Gamma (ip)\Gamma (-ip')}{\Gamma (ip+\ell +1)\Gamma (-ip'+\ell +1)}\int^\infty_{1/\epsilon}\frac{\dd r}{2\pi}e^{i(p-p')r}
        \notag\\
        &\quad\quad\quad\quad\quad\quad\quad
                                +\frac{\Gamma (-ip)\Gamma (ip')}{\Gamma (-ip+\ell +1)\Gamma (ip'+\ell +1)}\int^\infty_{1/\epsilon}\frac{\dd r}{2\pi}e^{-i(p-p')r}
                        \Biggr]
        \notag\\
        &\quad\quad
                =\frac{\Gamma (ip)\Gamma (-ip)}{\Gamma (ip+\ell +1)\Gamma (-ip+\ell +1)}
                        \delta (p-p')
        \,.\label{eq:scalar harmonics normalization}
}
We then have the normalized solution for the continuous mode $f^{p\ell}$ as
\al{
        f^{p\ell}(r_\rmJ )
                =&\sqrt{\frac{\Gamma (ip+\ell +1)\Gamma (-ip+\ell +1)}{\Gamma (ip)\Gamma (-ip)\sinh r_\rmJ}}
                        P^{-\ell -\frac{1}{2}}_{ip-\frac{1}{2}}(\cosh r_\rmJ )
        \,.\label{eq:f_pl}
}
Using the relation between coordinates eq.~\eqref{eq:r relation} we analytically
continue eq.~\eqref{eq:f^pl eq} to the region-C. 
We obtain the equation for the analytic-continued $f^{p\ell}$ as
\al{
        \biggl[
                -\frac{1}{\cosh^2 r}\frac{\dd}{\dd r}\left(\cosh^2 r\frac{\dd}{\dd r}\right)
                -\frac{\ell (\ell +1)}{\cosh^2 r}
        \biggr] f^{p\ell}(r)=(p^2+1)f^{p\ell}(r)
        \,.\label{eq:f^pl eq in C}
}
The analytic continuation of the eigenfunction to the region-C are given by
\al{
        &f^{p\ell}(r)
                =\sqrt{\frac{\Gamma (ip+\ell +1)\Gamma (-ip+\ell +1)}{i\Gamma (ip)\Gamma (-ip)\cosh r}}
                        P^{-\ell -\frac{1}{2}}_{ip-\frac{1}{2}}(i\sinh r )
        \,.\label{eq:normalized fpl}
}
We should note that the Wronskian relation for the Legendre functions
leads to the useful formula for the continuous modes~\cite{Garriga:1998he}:
\al{
        i\cosh^2 r
                \Biggl\{
                        \frac{\dd f^{p\ell}}{\dd r}\overline{f^{p\ell}}
                        -f^{p\ell}\frac{\dd\overline{f^{p\ell}}}{\dd r}
        \Biggr\}
                =\frac{2p}{\pi}\sinh (\pi p)
        \,.\label{eq:Wrinskian relation 1}
}

For the supercurvature mode with the imaginary wave number, the
normalization condition in eq.~\eqref{eq:normalization condition} is not
suitable. We then introduce the supercurvature eigenfunction
$f^{\Lambda\ell}$ with $\Lambda =ip$ as the solutions to
eq.~\eqref{eq:f^pl eq in C}. 
One possible choice of the normalization for 
the scalar harmonics for the discrete mode, $\mcY^{\Lambda\ell m}$\,,
are given by \cite{Garriga:1998he}
\al{
        &\mcY^{\Lambda\ell m}(r ,\Omega )=f^{\Lambda\ell}(r)Y_{\ell m}(\Omega )
        \,,\\
        &f^{\Lambda \ell}(r)
                =\sqrt{\frac{\Gamma (\Lambda +\ell +1)\Gamma (-\Lambda +\ell +1)}{2\cosh r}}
                        P^{-\ell -\frac{1}{2}}_{\Lambda -\frac{1}{2}}(i\sinh r )
        \,,\label{eq:f^Lambda ell def}
}
where we fix the normalization factor so that $\mcY^{\Lambda\ell m}$ are Klein-Gordon normalized in
the region-C, namely
\al{
        &i\cosh^2 r
                \int\dd\Omega\,
                \biggl\{
                        \left(\partial_r\mcY^{\Lambda\ell m}\right)\overline{\mcY^{\Lambda\ell' m'}}
                        -\mcY^{\Lambda\ell m}\left(\partial_r\overline{\mcY^{\Lambda\ell' m'}}\right)
                \biggr\}
        \notag\\
        &\quad
                =
        i\cosh^2 r
                \Biggl\{
                        \frac{\dd f^{\Lambda \ell}}{\dd r}\overline{f^{\Lambda\ell'}}
                        -f^{\Lambda\ell}\frac{\dd\overline{f^{\Lambda\ell'}}}{\dd r}
                \Biggr\}
                \int\dd\Omega Y_{\ell m}\overline{Y_{\ell' m'}}
                =\delta_{\ell\ell'}\delta_{mm'}
        \,.\label{eq:Wrinskian relation 2}
}

\subsection{Vector harmonics}
\label{sec:Vector harmonics}

The normalized (transverse) vector harmonics, $Y_i^{(\lambda )p\ell m}$
with $\lambda={\rm e}\,,{\rm o}$ being the parity of the harmonics\,,
are eigen functions of the Laplacian operator $\bar\nabla^2$ on the
3-dimensional hyperboloid in the region-J: 
\al{
        \bar\nabla^2 Y^{(\lambda )p\ell m}_i+(p^2+2)Y^{(\lambda )p\ell m}_i=0
        \,,\ \ \ 
        \bar\nabla^i Y_i^{(\lambda )p\ell m}=0
        \,.\label{eq:Y_i eq}
}
Apart from the normalization constant, the explicit expression for the vector harmonics
can be given by~\cite{Tomita:1982}
\al{
        &Y_r^{({\rm e})p\ell m}
                \propto\frac{1}{\sinh r_\rmJ}\mcP_{p\ell}(r_\rmJ )Y_{\ell m}(\Omega )
        \,,\\
        &Y_a^{({\rm e})p\ell m}
                \propto\frac{1}{\ell (\ell +1)}
                                                \frac{\dd}{\dd r_\rmJ}\Bigl[\sinh r_\rmJ\mcP_{p\ell}(r_\rmJ )\Bigr]
                        Y_{\ell m:a}(\Omega )
        \,,
}
for the even parity mode of the vector harmonics, and
\al{
        &Y_r^{({\rm o})p\ell m}
                =0
        \,,\ \ \ 
        Y_a^{({\rm o})p\ell m}
                \propto \sinh r_\rmJ\mcP_{p\ell}(r_\rmJ )
                        Y_{\ell m:b}(\Omega )\epsilon^b{}_a
        \,,
}
for the odd parity mode of the vector harmonics, where $\mcP_{p\ell}$
have been defined in eq.~\eqref{eq:f_pl 1}\,, the colon ( $:$ ) and
$\epsilon^a{}_b$ are the covariant derivative and the Levi-Civita symbol
on the unit two-sphere. On a two-dimensional spacetime, the Levi-Civita
symbol is a traceless antisymmetric rank-two tensor given by
\al{
        \epsilon_{ab}
                =\sin\theta
                        \left(
                                \begin{array}{cc}
                                0&1\\
                                -1&0\\
                                \end{array}                     
                        \right)
        \,.\label{eq:antisymmetric tensor}
}
One can verify the following properties:
\al{
        \epsilon_{ca}\epsilon^c{}_b=-\epsilon_{ac}\epsilon^c{}_b={\rm diag}(1,\sin^2\theta )
        \,,\ \ \  
        \epsilon_{ab:c}=0
        \,.
}
The normalization factor for the continuous mode ($p^2>0$) can be fixed
so that $Y_i^{(\lambda )p\ell m}$ satisfies
\al{
        &\int\dd r_\rmJ\dd\Omega\,\sinh^2 r_\rmJ
                \gamma^{ij}(r_\rmJ ,\Omega )
                Y^{(\lambda )p\ell m}_i(r_\rmJ ,\Omega )
                \overline{Y^{(\lambda' )p'\ell' m'}_j(r_\rmJ ,\Omega )}
        =\delta (p-p')\delta_{\lambda \lambda'}\delta_{\ell\ell'}\delta_{mm'}
        \,.\label{eq:normalization condition of vector}
}
As in the case of the scalar harmonics, the divergence contribution to
the integral comes only from the boundary of integration at 
$r_\rmJ =\pm\infty$\,. With a help of the asymptotic behavior of
$\mcP_{p\ell}$\,, we first evaluate the following integration for the
even parity mode of the vector harmonics: 
\al{
        &\int^\infty_0\dd r_\rmJ
                \biggl\{
                        \mcP_{p\ell}(r_\rmJ )\overline{\mcP_{p\ell}(r_\rmJ )}
                        +\frac{1}{\ell (\ell +1)}
                                \frac{\dd}{\dd r_\rmJ}\Bigl[\sinh r_\rmJ\mcP_{p\ell}(r_\rmJ )\Bigr]
                                \frac{\dd}{\dd r_\rmJ}\Bigl[\sinh r_\rmJ\overline{\mcP_{p\ell}(r_\rmJ )}\Bigr]
                \biggr\}
        \notag\\
        &\quad
                =\lim_{\epsilon\rightarrow 0}
                        \frac{pp'}{\ell (\ell +1)}
                                \Biggl[
                                        \frac{\Gamma (ip)\Gamma (-ip')}{\Gamma (1+\ell +ip)\Gamma (1+\ell -ip')}
                                        \int^{\infty}_{1/\epsilon}\frac{\dd r_\rmJ}{2\pi}e^{i(p-p')r_\rmJ}
        \notag\\
        &\quad\quad\quad\quad\quad\quad\quad\quad
                        +\frac{\Gamma (-ip)\Gamma (ip')}{\Gamma (1+\ell -ip)\Gamma (1+\ell +ip')}
                                \int^{\infty}_{1/\epsilon}\frac{\dd r_\rmJ}{2\pi}e^{-i(p-p')r_\rmJ}
                                \Biggr]
        \notag\\
        &\quad
                =\frac{p^2}{\ell (\ell +1)}
                        \frac{\Gamma (ip)\Gamma (-ip)}{\Gamma (ip+1+\ell )\Gamma (-ip+1+\ell )}
                        \delta (p-p')
        \,.
}
Hence, the explicit forms of the vector harmonics for the even parity mode 
are given by
\al{
        Y_r^{({\rm e})p\ell m}
                =&\frac{\sqrt{\ell (\ell +1)}}{p}\frac{1}{\sinh r_\rmJ}f^{p\ell}(r_\rmJ )Y_{\ell m}(\Omega )
        \,,\\
        Y_a^{({\rm e})p\ell m}
                =&\frac{1}{\sqrt{\ell (\ell +1)}p}
                        \frac{\dd}{\dd r_\rmJ}\Bigl[\sinh r_\rmJ\, f^{p\ell}(r_\rmJ )\Bigr] Y_{\ell m:a}(\Omega )
        \,.
}
where $f^{p\ell}(r_\rmJ )$ have been defined in eq.~\eqref{eq:f_pl}\,.
For the odd parity mode of the vector harmonics,
we can easily evaluate the normalization condition \eqref{eq:normalization condition of vector}
by using eq.~\eqref{eq:scalar harmonics normalization}\,.
We then obtain the explicit expression for the odd parity mode
of the vector harmonics as
\al{
        Y_r^{(o)p\ell m}=0
        \,,\ \ \ 
        Y_a^{(o)p\ell m}
                =\frac{1}{\sqrt{\ell (\ell +1)}}\sinh r_\rmJ f^{p\ell}(r_\rmJ )Y_{\ell m:b}(\Omega )\epsilon^b{}_a
        \,.
\label{eq:Odd normalization}
}
As with the case of the scalar harmonics eq.~\eqref{eq:Wrinskian relation 1}\,,
one can verify
\al{
        &i\cosh^2 r\int\dd\Omega\,\gamma^{ij}
                \biggl\{
                        \left(\pd_r Y_i^{(\lambda )p\ell m}\right)\overline{Y_j^{(\lambda' )p\ell' m'}}
                        -Y_i^{(\lambda )p\ell m}\left(\pd_r\overline{Y_j^{(\lambda' )p\ell' m'}}\right)
                \biggr\}
        \notag\\
        &\quad
                =i\cosh^2 r
                        \Biggl\{
                                \frac{\dd f^{p\ell}}{\dd r}\overline{f^{p\ell'}}
                                -f^{p\ell}\frac{\dd\overline{f^{p\ell'}}}{\dd r}
                        \Biggr\}
                        \delta_{\lambda\lambda'}\int\dd\Omega Y_{\ell m}\overline{Y_{\ell' m'}}
                =\frac{2p}{\pi}\sinh (\pi p)\,\delta_{\lambda\lambda'}\delta_{\ell\ell'}\delta_{mm'}
        \,.\label{eq:Wronskian relation for vector harmonics}
}
The vector harmonics for the supercurvature mode with $\Lambda =ip> 0$\,,
$\mcY_i^{(\lambda )\Lambda\ell m}$\,, are described in terms of
$f^{\Lambda\ell}$ defined in eq.~\eqref{eq:f^Lambda ell def} as
\al{
        &\mcY_r^{({\rm e})\Lambda\ell m}
                =\frac{\sqrt{\ell (\ell +1)}}{\Lambda}\frac{1}{\cosh r}f^{\Lambda\ell}(r)Y_{\ell m}(\Omega )
        \,,\\
        &\mcY_a^{({\rm e})\Lambda\ell m}
                =-\frac{1}{\sqrt{\ell (\ell +1)}\Lambda}\frac{\dd}{\dd r}
                        \Bigl[\cosh r f^{\Lambda\ell}(r)\Bigr] Y_{\ell m:a}(\Omega )
        \,,\label{eq:supercurvature even parity vector harmonics}
}
for the even parity,
\al{
        &\mcY_r^{({\rm o})\Lambda\ell m}=0
        \,,\ \ \ 
        \mcY_a^{({\rm o})\Lambda\ell m}
                =\frac{1}{\sqrt{\ell (\ell +1)}}i\cosh r f^{\Lambda\ell}(r)Y_{\ell m:b}(\Omega )\epsilon^b{}_a
        \,,\label{eq:supercurvature odd parity vector harmonics}
}
for the odd parity, where
$\mcY_i^{(\lambda )\Lambda\ell m}$ are Klein-Gordon normalized in the region-C:
\al{
        &i\cosh^2 r
                \int\dd\Omega\,
                        \gamma^{ij}
                        \biggl[
                                \left(\partial_r\mcY_i^{(\lambda )\Lambda\ell m}\right)\overline{\mcY_j^{(\lambda' )\Lambda\ell' m'}}
                                -\mcY_i^{(\lambda )\Lambda\ell m}\left(\partial_r\overline{\mcY_j^{(\lambda' )\Lambda\ell' m'}}\right)
                        \biggr]
        \notag\\
        &\quad
                =i\cosh^2 r
                        \Biggl\{
                                \frac{\dd f^{\Lambda\ell}}{\dd r}\overline{f^{\Lambda\ell'}}
                                -f^{\Lambda\ell}\frac{\dd\overline{f^{\Lambda\ell'}}}{\dd r}
                        \Biggr\}
                        \int\dd\Omega Y_{\ell m}\overline{Y_{\ell' m'}}
                =\delta_{\lambda\lambda'}\delta_{\ell\ell'}\delta_{mm'}
        \,,\label{eq:Wronskian relation for supercurvature vector harmonics}
}
where we have used eq.~\eqref{eq:Wrinskian relation 2}\,.

\section{Klein-Gordon norm}
\label{sec:KG norm}

In this section, we give the explicit expression for the Klein-Gordon norm
incorporating the non-dynamical field, following and extending
\cite{Mukohyama:1999kj}\,. Let us begin with the system of the
$m$-physical degrees of freedom $\phi^A$ ($A=1,\cdots ,m$) and
$n$-auxiliary variables $\varphi^\alpha$ 
($\alpha =m+1,\cdots ,m+n$). The discretized Lagrangian we will consider
here is given by 
\al{
        L=&\frac{1}{2}G_{AB}
                                \left(\dot\phi^A -f^A{}_C\phi^C -\tilde f^A{}_\alpha\varphi^\alpha\right)
                                \left(\dot\phi^B -f^B{}_D\phi^D -\tilde f^B{}_\beta\varphi^\beta\right)
        \notag\\
        &
                        -\frac{1}{2}V_{AB}\phi^A\phi^B -M_{A\alpha}\phi^A\varphi^\alpha 
                        -\frac{1}{2}\tilde V_{\alpha\beta}\varphi^\alpha\varphi^\beta
        \,.\label{eq:discretized Lagrangian}
}
We recast the Lagrangian in terms of the matrix description as
\al{
        L=&\frac{1}{2}
                                \left(\dot{\bm\phi}^{\rm T} -{\bm\phi}^{\rm T}{\bm f}^{\rm T} -{\bm\varphi}^{\rm T}\tilde{\bm f}^{\rm T}\right)
                                {\bm G}
                                \left(\dot{\bm\phi} -{\bm f}{\bm\phi} -\tilde{\bm f}{\bm\varphi}\right)
                        -\frac{1}{2}{\bm\phi}^{\rm T}{\bm V}{\bm\phi}
                        -{\bm\phi}^{\rm T}{\bm M}{\bm\varphi}
                        -\frac{1}{2}{\bm\varphi}^{\rm T}\tilde{\bm V}{\bm\varphi}
        \,.\label{eq:discretized Lagrangian}
}
The equation for $\varphi$ can be obtained by varying the Lagrangian \eqref{eq:discretized Lagrangian} as
\al{
        {\bm\varphi}
                =\left(\tilde{\bm f}^{\rm T}{\bm G}\tilde{\bm f}-\tilde{\bm V}\right)^{-1}
                        \Bigl[\tilde{\bm f}^{\rm T}{\bm G}\dot{\bm\phi}+\left({\bm M}^{\rm T}-\tilde{\bm f}^{\rm T}{\bm G}{\bm f}\right){\bm\phi}\Bigr]
        \,.\label{eq:phi^a}
}
Substituting the constraint equation \eqref{eq:phi^a} into the Lagrangian \eqref{eq:discretized Lagrangian}\,,
after lengthy calculation, we obtain the reduced Lagrangian as
\al{
        L=&\frac{1}{2}
                                \left(\dot{\bm\phi}^{\rm T} -{\bm\phi}^{\rm T}{\bm f}_{\rm eff}^{\rm T}\right)
                                {\bm G}_{\rm eff}
                                \left(\dot{\bm\phi} -{\bm f}_{\rm eff}{\bm\phi}\right)
                                -\frac{1}{2}{\bm\phi}^{\rm T}{\bm V}_{\rm eff}{\bm\phi}
        \,,\label{eq:reduced Lagrangian}
}
where
\al{
        &{\bm f}_{\rm eff}
                ={\bm f}-\tilde{\bm f}\tilde{\bm V}^{-1}{\bm M}^{\rm T}
        \,,\\
        &{\bm G}_{\rm eff}
                ={\bm G}-{\bm G}\tilde{\bm f}\left(\tilde{\bm f}^{\rm T}{\bm G}\tilde{\bm f}-\tilde{\bm V}\right)^{-1}\tilde{\bm f}^{\rm T}{\bm G}
        \,,\\
        &{\bm V}_{\rm eff}
                ={\bm V}-{\bm f}^{\rm T}{\bm G}{\bm f}
                                        +\left({\bm M}-{\bm f}^{\rm T}{\bm G}\tilde{\bm f}\right)
                                                \left(\tilde{\bm f}^{\rm T}{\bm G}\tilde{\bm f}-\tilde{\bm V}\right)^{-1}
                                                \left({\bm M}^{\rm T}-\tilde{\bm f}^{\rm T}{\bm G}{\bm f}\right)
                +{\bm f}_{\rm eff}^{\rm T}{\bm G}_{\rm eff}{\bm f}_{\rm eff}
        \,.
}
We have used the useful formula as
\al{
        {\bm G}_{\rm eff}{\bm f}_{\rm eff}
                ={\bm G}{\bm f}+{\bm G}\tilde{\bm f}\left(\tilde{\bm f}^{\rm T}{\bm G}\tilde{\bm f}-\tilde{\bm V}\right)^{-1}\left({\bm M}^{\rm T}-\tilde{\bm f}^{\rm T}{\bm G}{\bm f}\right)
        \,.\label{eq:useful formula}
}
We now promote ${\bm\phi}$ to operators $\hat{\bm\phi}$\,, and 
expand $\hat{\bm\phi}$ by mode functions $\{{\bm\phi}_\mcN\,,\overline{{\bm\phi}_\mcN}\}$\,,
which is expressed as
\al{
        \hat{\bm\phi}=\sum_\mcN\left( a_\mcN{\bm\phi}_\mcN +a_\mcN^\dagger\overline{{\bm\phi}_\mcN}\right)
        \,,
}
where $\hat a_\mcN$ and $\hat a_\mcN^\dagger$ are the annihilation and creation operators, respectively,
we assume that $\{{\bm\phi}_\mcN\,,\overline{{\bm\phi}_\mcN}\}$ forms a complete set of linear independent
solutions of the equation of motion.
The quantum fluctuations of the field are described by the vacuum state\,, which is annihilated by
the annihilation operator, $\hat a_\mcN |0\rangle =0$\,.
According to \cite{Mukohyama:1999kj}\,, the discretized KG norm for the reduced Lagrangian \eqref{eq:reduced Lagrangian} 
can be defined by
\al{
        \left({\bm\phi}_\mcN ,{\bm\phi}_\mcM\right)_{\rm KG}
                =-i\biggl\{
                        {\bm\phi}^{\rm T}_\mcN{\bm G}_{\rm eff}
                        \left(\overline{\dot{\bm\phi}_\mcM} -{\bm f}_{\rm eff}\overline{{\bm\phi}_\mcM}\right)
                        -\left(\dot{\bm\phi}^{\rm T}_\mcN -{\bm\phi}^{\rm T}_\mcN{\bm f}_{\rm eff}^{\rm T}\right)
                        {\bm G}_{\rm eff}\overline{{\bm\phi}_\mcM}
                \biggr\}
        \,.
}
With the help of eqs.~\eqref{eq:phi^a} and \eqref{eq:useful formula}\,,
the KG norm for the physical degrees of freedom ${\bm\phi}$ can be
reduced to the simple form in terms of the auxiliary variables
${\bm\varphi}$ as
\al{
        \left({\bm\phi}_\mcN ,{\bm\phi}_\mcM\right)_{\rm KG}
                =-i\biggl\{
                        {\bm\phi}^{\rm T}_\mcN{\bm G}
                        \left(\overline{\dot{\bm\phi}_\mcM} -{\bm f}\overline{{\bm\phi}_\mcM} -\tilde{\bm f}\overline{{\bm\varphi}_\mcM}\right)
                        -\left(\dot{\bm\phi}^{\rm T}_\mcN -{\bm\phi}^{\rm T}_\mcN{\bm f}^{\rm T}-{\bm\varphi}_\mcN^{\rm T}\tilde{\bm f}^{\rm T}\right)
                        {\bm G}\overline{{\bm\phi}_\mcM}
                \biggr\}
        \,,
}
that is,
\al{
        \left({\bm\phi}_\mcN ,{\bm\phi}_\mcM\right)_{\rm KG}
                =-iG_{AB}\biggl\{
                        \phi_\mcN^A
                        \left(\overline{\dot\phi_\mcM^B} -f^B{}_D\overline{\phi_\mcM^D} -\tilde f^B{}_\beta\overline{\varphi_\mcM^\beta}\right)
                        -\left(\dot\phi_\mcN^A -f^A{}_C\phi_\mcN^C -\tilde f^A{}_\alpha\varphi_\mcN^\alpha\right)
                        \overline{\phi_\mcM^B}
                \biggr\}
        \,,
}


\end{document}